\documentclass{bmcart}

\usepackage{amsthm,amsmath}
\usepackage{amssymb}
\usepackage{graphicx}
\usepackage{url}
\usepackage[utf8]{inputenc} 

\startlocaldefs
\endlocaldefs

\begin{document}

\begin{frontmatter}

\newtheorem{theorem}{Theorem}
\newtheorem{corollary}{Corollary}
\newtheorem{lemma}{Lemma}
\newtheorem{proposition}{Proposition}
\newtheorem{definition}{Definition}

\newtheorem{example}{Example}
\newtheorem{remark}{Remark}

\begin{fmbox}
\dochead{Research}


\title{The energy-spectrum of bicompatible sequences}


\author[
   addressref={aff1},                   
   email={fwh3zc@virginia.edu}   
]{\inits{FWH}\fnm{Fenix W.} \snm{Huang}}
\author[
   addressref={aff1,aff2},
   email={ChrisBarrett@virginia.edu}
]{\inits{CLB}\fnm{Christopher L.} \snm{Barrett}}
\author[
   addressref={aff1,aff3}, 
   corref={aff3},   
   email={duckcr@gmail.com}
]{\inits{CMR}\fnm{Christian M.} \snm{Reidys}}


\address[id=aff1]{
  \orgname{Biocomplexity Institute \& Initiative, University of Virginia}, 
  \street{995 Research Park Blvd., Suite 400},                    %
  \postcode{VA 22911},                                
  \city{Charlottesville},                              
  \cny{USA}                                    
}
\address[id=aff2]{
  \orgname{Department of Computer Science, University of Virginia}, 
  \postcode{VA 22904},                                
  \city{Charlottesville},                              
  \cny{USA}                                    
}
\address[id=aff3]{
  \orgname{Department of Mathematics, University of Virginia}, 
  \postcode{VA 22904},                                
  \city{Charlottesville},                              
  \cny{USA}                                    
}



\end{fmbox}


\begin{abstractbox}

\begin{abstract} 
\parttitle{Background} 
Genotype-phenotype maps provide a meaningful filtration of sequence space
and RNA secondary structures are particular such phenotypes. Compatible sequences i.e.~sequences that
satisfy the base pairing constraints of a given RNA structure play an important role in the context of
neutral networks and inverse folding.
Sequences satisfying the constraints of two structures simultaneously are called bicompatible and
phenotypic change, induced by erroneously replicating populations of RNA sequences, is closely connected
to bicompatibility. Furthermore, bicompatible sequences are relevant for riboswitch sequences, beacons of
evolution, realizing two distinct phenotypes.
\parttitle{Results}  
We present a full loop energy model Boltzmann sampler of bicompatible sequences for pairs of structures.
The novel dynamic programming algorithm is based on a topological framework encapsulating the relations between loops. 
We utilize our sequence sampler to study the energy spectra and density of bicompatible sequences, the rankings of the structures and key properties for evolutionary transitions. 
\parttitle{Conclusion} 
Our analysis of riboswitch sequences shows that key properties of bicompatible sequences depend on the particular pair
of structures. While there always exist bicompatible sequences for random structure pairs, they are less
suited to facilitate transitions.
We show that native riboswitch sequences exhibit a distinct signature with regards to
the ranking of their two phenotypes relative to the minimum free energy, suggesting a new criterion
for identifying native sequences and sequences subjected to evolutionary pressure.\\
Our free software is available at: \url{https://github.com/FenixHuang667/Bifold}
\end{abstract}




\end{abstractbox}
%

\end{frontmatter}




\section*{Background}\label{S:intro}

RNA evolution has been studied extensively in the framework of theoretical evolutionary optimization,
center staging the genotype-phenotype mapping from RNA sequences to their structures
\cite{Fontana:93,Gruner:96,Gruner:96b,Schuster:94,Reidys:97,Fontana:98,Schuster:97}.
RNA secondary structures are particular such phenotypes. They are contact structures, that can be represented as 
diagrams with noncrossing arcs in the upper half plane. In addition, they correspond
to tree structures and are consequently well suited for a variety of recursions and dynamic programming (DP)
routines, based on the lengths of contiguous subsequences.

Schuster \cite{Schuster:94} realized that genotype-phenotype mappings provide a natural filtration of
the sequence space
by means of considering sequences ``equivalent'' if they fold into the same minimum free energy (mfe)
secondary structure. This perspective naturally leads to analyzing the induced subgraphs of preimages
in the sequence space and to the concept of neutral networks of RNA secondary structures \cite{Reidys:97}.
These in turn allow studying well-known evolutionary theories, such as Motoo Kimura's neutral theory of
evolution.

A plethora of work has been done on the diffusion-like process of sequences searching for an optimal
structure, ranging from simulation-based studies \cite{Kimura:68} to the mathematical analysis of the cluster-size
distribution depending on the structure of the neutral net \cite{Reidys:97}. These studies have shown that
connectivity and density of neutral networks are of central importance for the understanding of how
sequences evolve.

One prominent phenomenon is that of spontaneous, rapid transitions of evolving populations of RNA sequences
from one structure to another--even in absence of fitness advantages \cite{Fontana:98, Forst:95}.
Despite the Intersection theorem \cite{Reidys:97}, guaranteeing the existence of bicompatible sequences for
\emph{any} two RNA secondary structures, transitions between neutral networks are only observed for
certain structure pairs. Weber \cite{Forst:95} showed that in the course of a phenotypic transition an evolving
population of RNA sequences tunnels through bicompatible sequences. These sequences represent a gateway
between different phenotypes.

\begin{figure}[h!]
  \begin{center}
\includegraphics[width=0.9\columnwidth]{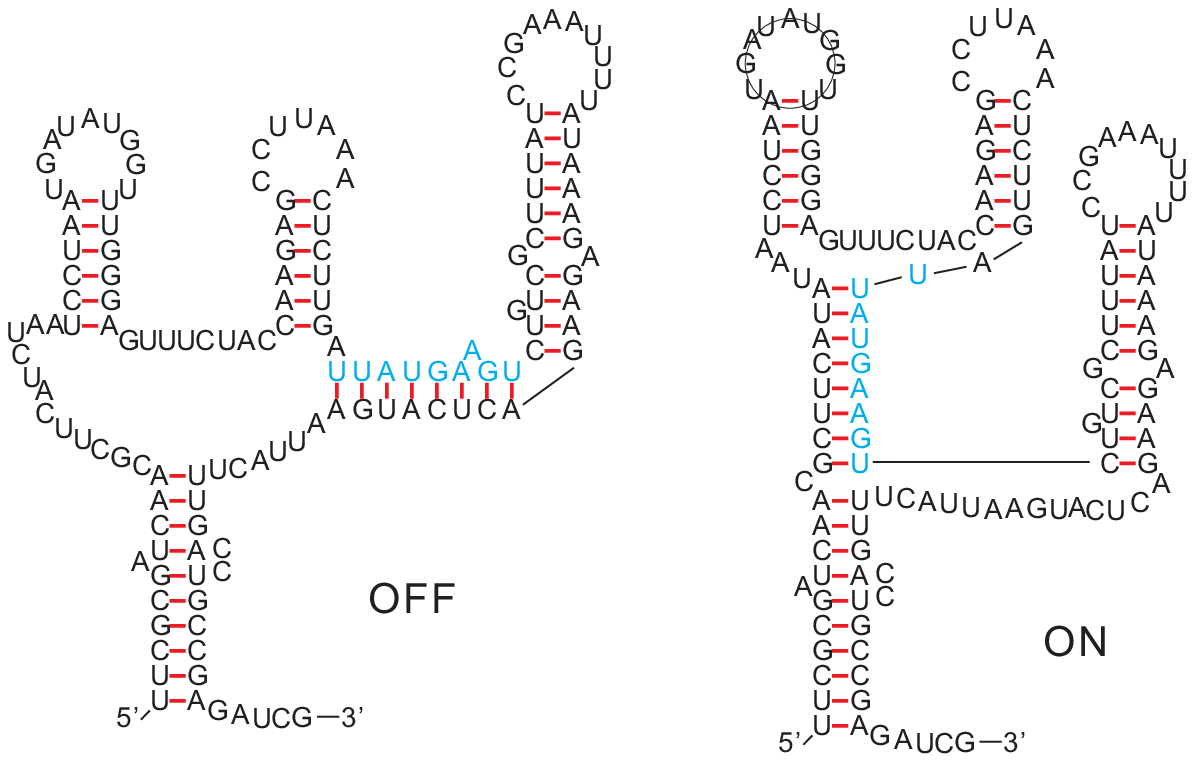}
\end{center}
  \caption{\csentence{Riboswitch.}
      Alternative structures of the Adenine riboswitch \cite{add} and its switching sequence
 (blue), involved two respective helices}
 \label{F:riboswitch}
 \end{figure}

Bicompatible sequences play furthermore a prominent role in the analysis of RNA riboswitches
\cite{Garst:11}. Riboswitch sequences express two distinct structures, each of
which appearing in a specific biophysical contexts, see Figure~\ref{F:riboswitch}.
Both structures are typically thermodynamically suboptimal and exhibit a large base pair distance
\cite{Serganov:13}. Specific mechanisms are observed, most prominently that of the existence of a
\emph{switching sequence}, a contiguous subsequence that engages for each respective structure
in a unique fashion. The two structures are mutually exclusive, since bases of the switching
sequence pair downstream in one and upstream in the other configuration \cite{Garst:11}.


The studies of bicompatible sequences of a pair of secondary structures, or in more general cases where a sequence satisfies multiple structure constraints, are motivated by the computational design of RNA sequences  \cite{Flamm:01, lyngso2012frnakenstein,taneda2015multi, Hammer:17, Hammer:19}. 
Early methods such as \cite{Hofacker:94a, Flamm:01} design a sequence by simulation approaches. These entail considering an objective function that involves the energy contribution of 
multiple target structures on a common sequence. They then replicate a sequence by introducing a single random mutation, and a sequence survives if it improve the performance of the 
designated objective function. 
Later developments such as \cite{lyngso2012frnakenstein,taneda2015multi} generate sequences satisfying multiple structure constraints based on a multi-objective genetic algorithm. 
Recently, new approaches such as \cite{Levin:12, Reinharz:13, Clote:16, Barrett:17}, design sequences on a single secondary structure using a Boltzmann sequence sampler. The sampler is based on computing a ``dual'' partition function of sequences for a fixed structure. This provides key information for sequences compatible with a fixed secondary structure. Dual here is meant with respect to the partition function of secondary structures introduced by McCaskill \cite{McCaskill:90}. 
For multiple structures, \cite{Hammer:17} presents an algorithm to sample sequences with multiple structure constraints with uniform probability. Further development in \cite{Hammer:19} 
allows one to consider a simplified loop-based energy model, which enables Boltzmann sampling sequences with multiple structure constraints.

\cite{Hammer:19} employs a dynamic programming routine (DP) to compute the dual partition functions for a fixed decomposition of a bistructure. The sampler 
use a hyper-graph model to describe the intersections of loops. Assuming there is a tree decomposition of the hyper-graph, the time complexity of computing the partition function 
is $O(4^{w+1} n)$ where $w$ is the tree-width of the given decomposition and $n$ is the length of the given bistructure. Therefore, computing the partition function is 
fixed-parameter tractable (FPT). However, obtaining a tree decomposition of the hyper-graph that minimizes the tree-width is \#P-hard \cite{Hammer:19}. 
For a fixed natural number $k$, checking whether a tree decomposition with tree-width $\le k$ exists, can be performed in linear time \cite{bodlaender1996linear} and if such a decomposition exists, it can be computed in linear time. However, no relation between a given bistructure and the parameter $k$ is given in \cite{Hammer:19}. Furthermore, although the above check can be performed relatively quickly, the tree decomposition produced can have tree-width $<k$, and it is not clear how large of a gap exists between the tree-width and $k$. 

In this paper, we focus our analysis on sequences that are compatible with a given pair of secondary structures. We are interested in understanding the phenotypic transition that occurs between the two. Our framework straightforwardly generalizes the case of sequences that are subject to constraints of multiple structures. Our two main objectives are: a) providing a novel topological framework facilitating deeper understanding of the algorithmic complexity of the dual sampler, and b) employing a Boltzmann sampler for bicompatible sequences based on Turner's energy model \cite{Mathews:99} in order to investigate intrinsic features of riboswitches. That is, we are dealing with the full loop energy model, considering energy contributions from various loop types like hairpin-loops, interior-loops, and multi-loops. The features investigated here cannot be inferred by sequence alignment methods. 

As for a), we base our framework on the homological analysis of bistructures \cite{Bura:19}, where a loop is a set of vertices, and intersections are encoded via a simplicial complex. 
A loop is viewed as a $0$-simplex, and $d$ loops having nontrivial intersection as $(d-1)$-simplices. The $0$-simplices are hyper-edges of the model discussed in \cite{Hammer:19}, while $d$-simplices for $d>0$ capture additional information not present in the hyper-graph model. The collection of all $d$-simplices forms a simplicial complex called the {\it loop nerve}. It gives rise to a topological space providing a quantification of algorithmic complexity via homology groups. It is shown in \cite{Bura:19} that only the second homology group is relevant. This gives a natural classification of a bistructure $B$ by its rank, $r_2(B)$. Based on this classification, the topological space is a ribbon tree modulo certain contraction of its spheres.
The knowledge about this space allows us to derive a decomposition of bistructures.
We shall derive an optimal algorithm to compute the partition function of bistructures having $r_2(B) = 0$, i.e., we can obtain in this case a tree decomposition of the hyper-graph in~\cite{Hammer:19} with minimal tree-width. For $r_2(B) > 0$, designing an optimal algorithm is \#P-hard, and the root of the algorithmic complexity resides in the processing of the spheres. The latter are given a combinatorial interpretation, namely, as the crossing components in the arc diagram. We are developing an approach to resolve a sphere by mapping it to a known NP-problem like, for instance the traveling salesman problem (TSP). Via such a mapping, efficient approximation algorithms can be employed resolving the spheres \cite{rego2011traveling}.

As for b) it is well-known \cite{tadrowski2018phenotypic, oostra2018strong} that even identical genotypes can lead to many different phenotypes in response to environmental changes . 
It is thus important to understand the phenotypic accessibility through bicompatible sequences. To this end, 
we implement a sequence sampler for bistructures based on the Turner model \cite{Mathews:99} and provide a detailed analysis of multiple riboswitch sequences on multiple levels. 
We first study various types of free energies of bicompatible sequences in relation to those of compatible ones. 
We show that for the two alternative structures $R$ and $S$ of a riboswitch, modifying an $R$-compatible sequence into a $(R,S)$-bicompatible sequence
can be done without affecting the free energy with respect to $R$. This is also simultaneously holds for $S$, and is not observed for bicompatible sequences for two random structures.  
The result shows that riboswitches exhibit a clearly distinguishable signal from that of random sequences with respect to accessibility. 
We further analyze how the structure pairs rank within the partition function by comparing riboswitch sequences with random sequences. 
From the rank analysis we can conclude that the relative rank of the native riboswitch sequences is distinctively higher than the relative rank of the sampled sequences. This indicates that the native sequence exhibits an evolved thermodynamic stability with respect to the pair of structures $(R,S)$. Finally, we investigate the density of bicompatible sequences within the sets of $R$ and $S$ compatible sequences respectively. The density of riboswitch sequences is distinctively different from that of random structure pairs. This indicates that the two alternative structures of a riboswitch are more evolutionary accessible by sequences than the random structure pairs.


\section*{Materials and Methods}

\subsection*{Bistructures}
We present an RNA secondary structure as a diagram, a graph whose vertices are drawn on a horizontal
line and the Watson-Crick as well as Wobble base pairs are drawn as arcs in the upper half plane. 
The vertices are labeled by $V=\{1,2,\ldots, n\}$ from left to right, representing the nucleotides.
The linear order of the vertices indicates the direction of the backbone from $5'$-end to $3'$-end.
Furthermore, each vertex can be paired with at most one other vertex by an arc drawn in the upper
half-plane. An arc, $(i,j)$, represents the base pair between the $i$th and $j$th nucleotides. Two
arcs $(i,j)$ and $(r,s)$ are called crossing if and only if  $i<r$ and $i<r<j<s$ holds. An RNA structure
is called a secondary structure, if it does not contain any crossing arcs. Furthermore,
the arcs of a secondary structure can be endowed with the partial order: $(r,s) \prec (i,j)$ if and only
if $i<r<s<j$. We shall introduce two ``formal'' vertices associated with positions $0$ and $(n+1)$, respectively
and add the formal arc $(0, n + 1)$, referred to as the {\it rainbow}. An \emph{interval}, $[i,j]$, is the
set of vertices $\{i,i+1,\ldots,j-1,j\}$. 

In a loop-based energy model  \cite{Mathews:99, Mathews:04}, arcs and unpaired vertices are organized
in \emph{loops} contributing to the energy.
A loop, $L$, is a subset of vertices, represented as a disjoint
union of $S$-intervals, $L=\dot\bigcup_{i=1}^k [a_i,b_i]$, such that $(a_1,b_k)$ and
$(b_i,a_{i+1})$, for $1\leq i\leq k-1$, are arcs (including the rainbow arc $(0, n+1)$) and where any other interval-vertices are
unpaired. It can be represented by a maximal arc $(a_1,b_k)$ with respect to the partial order $\prec$. Given a loop, this maximal arc is unique, whence a loop can be represented by
$L_{(a_1, b_k)}$. In particular, the rainbow arc, $(0,n+1)$, represents an exterior loop, that is not nested in any arc in the arc diagram.
Furthermore, each non-rainbow arc appears in exactly two loops, being maximal for exactly one of them. 
Loops correspond to the boundary components of the secondary structure viewed as a fatgraph
\cite{Penner:10}. In the following, we shall identify loops with their sets of vertices.

Given two secondary structures, $R$ and $S$, having the same vertex set $V=\{1,\dots,n\}$, we draw the vertices on a
horizontal line, the arcs of $R$ in the upper and the arcs of $S$ in the lower half-plane. 
We refer to this arc diagram as a {\it bistructure}, $B(R,S)$. 
Here we shall distinguish the $R$-arcs from the $S$-arcs even though they might have the exact same endpoints. For example an arc $(i,j)$ in $R$ is denoted by  
$(i,j)_R$ and an arc $(i,j)$ in $S$ is denoted by $(i,j)_S$. 
In a loop-based model, the $R$-loops and the $S$-loops are distinct since their represented $R$-arcs from the $S$-arcs are distinct.
Hence, a bistructure $B(R,S)$ can be considered as the set of loops 
$B = \{L_{p_i} \mid p_i \in B(R,S), 1\le i \le m\}$, where $p_i$, $1\le i \le m$, is an arc in $B(R,S)$.

\begin{figure}[ht]
\begin{center}
\includegraphics[width=0.9\columnwidth]{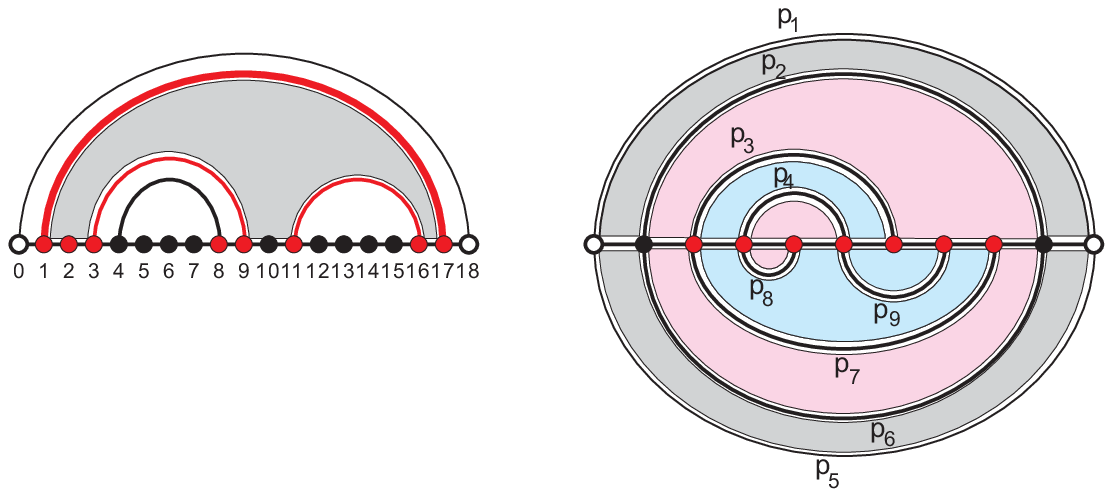}
\end{center}
\caption{\small (LHS) A single secondary structure:  a loop (gray) contains a distinguished maximal arc $(1, 17)$.   
(RHS) A bistructure $B=\{L_1, \ldots, L_9\}$. $X = \{L_3, L_7, L_9\}$ (blue) is an irreducible substructure of $B$ with its complement $\overline{X} = \{L_1, L_2, L_4, L_5, L_6, L_8\}$. 
We mark the exposed vertices $E^X = V^X \cap V^{\overline{X}}$ in red. The closure of $X$ is given by $\tilde{X} = \{L_2, L_3, L_4, L_6, L_7, L_8, L_9\}$. 
}
\label{F:bistructure}
\end{figure}

A {\it substructure} of $B$, denoted by $B'$, is a subset of loops where $B' \subseteq B$. The vertex set of $B'$, denoted by $V^{B'}$, is the union of vertices in loops that are contained in 
$B'$. The complement of $B'$, $\overline{B'} = B \setminus B'$, with its vertex set $\overline{V^{B'}}$, see Figure~\ref{F:bistructure}.
Accordingly, we have (a) $V^{B'}\cup \overline{V^{B'}}$ contains all vertices
in $B(R,S)$ and (b) $V^{B'}\cap \overline{V^{B'}}$ is not necessarily empty, since paired and
unpaired vertices can be contained in the intersection of the $B'$- and $\overline{B'}$-loops. 
Furthermore, for a given substructure $X=\{L_1,\cdots,L_k\}$, we define the boundary of $X$ by $X^C = \{ P_j\in \overline{X}| \exists L_i\in X, P_j\cap L_i\ne\varnothing\}$. I.e., $X^C$ is the set of all loops in the complement of $X$ that have nontrivial intersection with $X$. We call $\tilde{X} = X\cup X^C$ the {\it closure} of $X$.
A substructure is called {\it reducible} if the loop set can be bi-partitioned into two sets of loops 
$X_1 = \{L_{i_1}, \ldots, L_{i_m}\}$ and $X_2 = \{L_{j_1}, \ldots, L_{j_n}\}$, such that $L_{i_t} \cap L_{j_s} =\varnothing$, $\forall 1\le t \le m$, $1\le s \le n$, otherwise we call $X$ {\it irreducible}.

The intersection $E^{B'}=V^{B'}\cap \overline{V^{B'}}$ is called the set of \emph{exposed} vertices of $B'$. The exposed vertices are key elements in computing the 
partition function of a bistructure, since the vertices are contained in multiple loops and their nucleotide information needs to be remembered 
until the energies of the loops containing the exposed vertices are calculated. 

\subsection*{Partition function and Boltzmann sampler}\label{S:pf}

We first recall the notion of a partition function for sequences that are compatible to a single structure $R$ \cite{McCaskill:90}. 
$$
Q(R) = \sum_{\sigma \in \mathbb{C}_n(R)} e^{-\frac{\eta(\sigma, R)}{KT}}. 
$$
Here $\mathbb{C}_n(R)$ denotes the set of $R$-compatible sequences while $\eta(\sigma, R)$ is the energy of the sequence-structure pair $(\sigma, R)$. Lastly, $K$ is the Boltzmann constant and
$T$ the temperature. In Turner's model \cite{Mathews:99, Mathews:04}, $\eta(\sigma, R) = \sum _{L\in R} \eta(\sigma, L)$, where $L$ is a loop contained in the secondary structure $R$. 
The energy of a loop $L$ is a function of its type and of the nucleotides associated to the arcs and the unpaired
bases it contains. In practice, the energy computation takes into account a maximum of two specific
arcs and four unpaired vertices, as well as the {\it number} of arcs and the number of unpaired bases.

For a bistructure $B(R,S)$ and a sequence $\sigma$, we set $\eta(\sigma, B(R,S)) = \frac{1}{2}(\eta(\sigma,R) + \eta(\sigma,S))$. 
Then we define the partition function of sequences bicompatible to $R$ and $S$ by 
$$
Q(R ,S) = \sum_{\sigma \in \mathbb{C}_n(R,S)} e^{-\frac{\eta(\sigma, B(R,S))}{KT}},
$$
where $\mathbb{C}_n(R,S)$ denotes the set of bicompatible sequences to both $R$ and $S$. 


A decomposition of $B$ is a block sequential loop removal of the bistructure. 
Let us first illustrate the computation of $Q(R,S)$ when a specific decomposition is given.
Suppose $X = \{L_1,\ldots, L_k\}$ is a substructure of $B(R,S)$ with vertex set $V$,  and exposed vertex set $E^X$. $\overline{X} =B \setminus X$ denotes the complement of $X$. 
Let $\sigma_X = (\sigma_v)_v$ denote a subsequence with $v\in V$, $\sigma_v \in \{{\bf A, U, G,C}\}$. Then we can compute the energy $\eta(\sigma_X, X)$ since the nucleotide information of the vertices contained in $V$ is specified. 
Let further $\tau_X = (\tau_v)_v$ be a subsequence where $v\in E^X$, $\tau_v \in \{{\bf A, U, G,C}\}$. Clearly, $\tau_X \subseteq \sigma_X$. For $\ell = |V|$, we define a partition function for $X$ that is parameterized by $\tau_X$
$$
Q(X, \tau_X) = \sum_{\sigma_X \in \mathcal{Q}^4_{\ell}} e^{-\frac{\eta(\sigma_X, X)}{KT}}. 
$$

By definition, if $X$ is an irreducible substructure, then removing a loop $L$ from $X$ produces a set of irreducible substructures $X_1,\ldots, X_k$. We investigate how the exposed vertex set evolves with a loop removal. To this end let $x\in E^X$ be an exposed vertex. If $x\in L$, then either a) $\nexists L'\in X, L'\ne L$ such that $x\in L'$, or , b) at least one such $L'$ loop exists. 
In the first case a), we have $x$ is no longer exposed, while in the second case b), we have $x\in E^{X_i}$ for some $1\le i\le k$. 
Finally, if $x\notin L$ to begin with, then we  have $x\in E^{X_i}$ for some $1\le i\le k$ after removing $L$ form $X$.

Let $\tau_X$ denote a fixed subsequence over $E^X$, $\tau_{X_i}$ a subsequence over $E^{X_i}$, $1\le i \le k$, and $\sigma_L$ a subsequence over the loop $L$.  
We consider all possible subsequences $(\sigma_v)_v$ where $\sigma_v \in \{{\bf A, U, G,C}\}$, $v\in \left(L \cup_{i=1}^k E^{X_i} \right) \setminus E^X$. 
Then, the partition function $Q(X, \tau_X)$ can be computed recursively by
\begin{equation}\label{E:recursion}
Q(X, \tau_X) = \sum_{(\sigma_v)_v} e^{-\frac{\eta((\sigma_L, L)}{KT}} \prod_i^k Q(X_i, \tau_{X_i}). 
\end{equation}
For a given decomposition, the terms $Q(X_i, \tau_{X_i})$, for $1\le i \le k$, can be computed in parallel.

For a fixed decomposition $D$ of $B(R,S)$, the time complexity of computing eq.~\ref{E:recursion} is given by the maximum number of $ |L \cup_{i=1}^k E^{X_i}|$ in all recursion steps. We denote this number by $\kappa_D(B)$. 
Let further $\kappa(B) = \min_D \kappa_D(B)$, i.e. the minimum time complexity over all possible decompositions $D$.  
Clearly, $\kappa(B)$ depends only on the bistructure $B$. For a fixed $\kappa(B)$, we can implement a dynamic programming (DP) routine to
compute $Q(R,S)$ recursively \cite{Hammer:19}. The time complexity of the algorithm is  $O(4^{\kappa(B)} n)$ since for every $\sigma_v$, $v\in L \cup_{i=1}^k E^{X_i}$, we have four nucleotides choices 
${\bf A, U, G,C}$. 
Here $n$ is the length of the given bistructure. 
Therefore, the algorithm to compute $Q(R,S)$ is a (FPT) algorithm. This is in accordance with the result reported in \cite{Hammer:19}, and $\kappa(B) = w+1$, where $w$ is the tree-width of a fixed decomposition tree
of the hyper-graph discussed  in \cite{Hammer:19}. 

When $Q(R,S)$ is computed, we can Boltzmann sample RNA sequences following the classical stochastic backtracking method introduced by \cite{Ding:03}, which is of linear time complexity. 
Given an irreducible substructure $X$ that is decomposed into a loop $L$ and a set of irreducible substructures $X_1, \ldots, X_k$. Assume the nucleotides in $X_i$ are sampled, then 
with a fixed subsequence $\tau_X$ over the exposed vertex set $E^X$, the subsequence $(\sigma_v)_v$, $v\in L \setminus \cup_i E^{X_i}$ is sampled with probability
$$
\frac{ e^{-\frac{\eta((\sigma_v)_v, L)}{KT}} \prod_i^k Q(X_i, \tau_{X_i}) }{ Q(X, \tau_X) }.
$$
Multiplying all inside probabilities of each iteration, we conclude that a sequence is
sampled with probability 
$\mathbb{P}(\sigma) = e^{-\frac{\eta(\sigma, B(R,S))}{KT}}/Q(R,S)$.

\subsection*{The topology of bistructure} \label{S:strategy}

For a given decomposition of a bistructure, the time complexity of computing its partition function is polynomial. 
The question is, how to design a decomposition for a bistructure that minimizes $\kappa_D(B)$. 
\cite{Hammer:19} argues that this design problem is \#-P hard, however falling short of providing methods to construct such a decomposition. 
For a given natural number $k$ fixed, although checking whether a tree decomposition with tree-width $\le k$ exists, as well as constructing such a decomposition, can be 
performed in linear time \cite{bodlaender1996linear}, it is however not clear, which $k$ is optimal. 

Loop intersections are studied in \cite{Bura:19} via the loop nerve, interpreting a loop in a bistructure $B(R,S)$ as an abstract $0$-simplex. 
If $d$ loops have nonempty mutual intersection, they are represented by a $(d-1)$-simplex. The collection of all $d\ge 0$-simplices forms a simplicial complex. A loop removal is tantamount to deleting the corresponding $0$-simplex as well as all higher dimensional simplices that contain it. We shall show that understanding the structure of the topological space provides insight into designing an optimal decomposition. 

\begin{figure}[ht]
\begin{center}
\includegraphics[width=0.9\columnwidth]{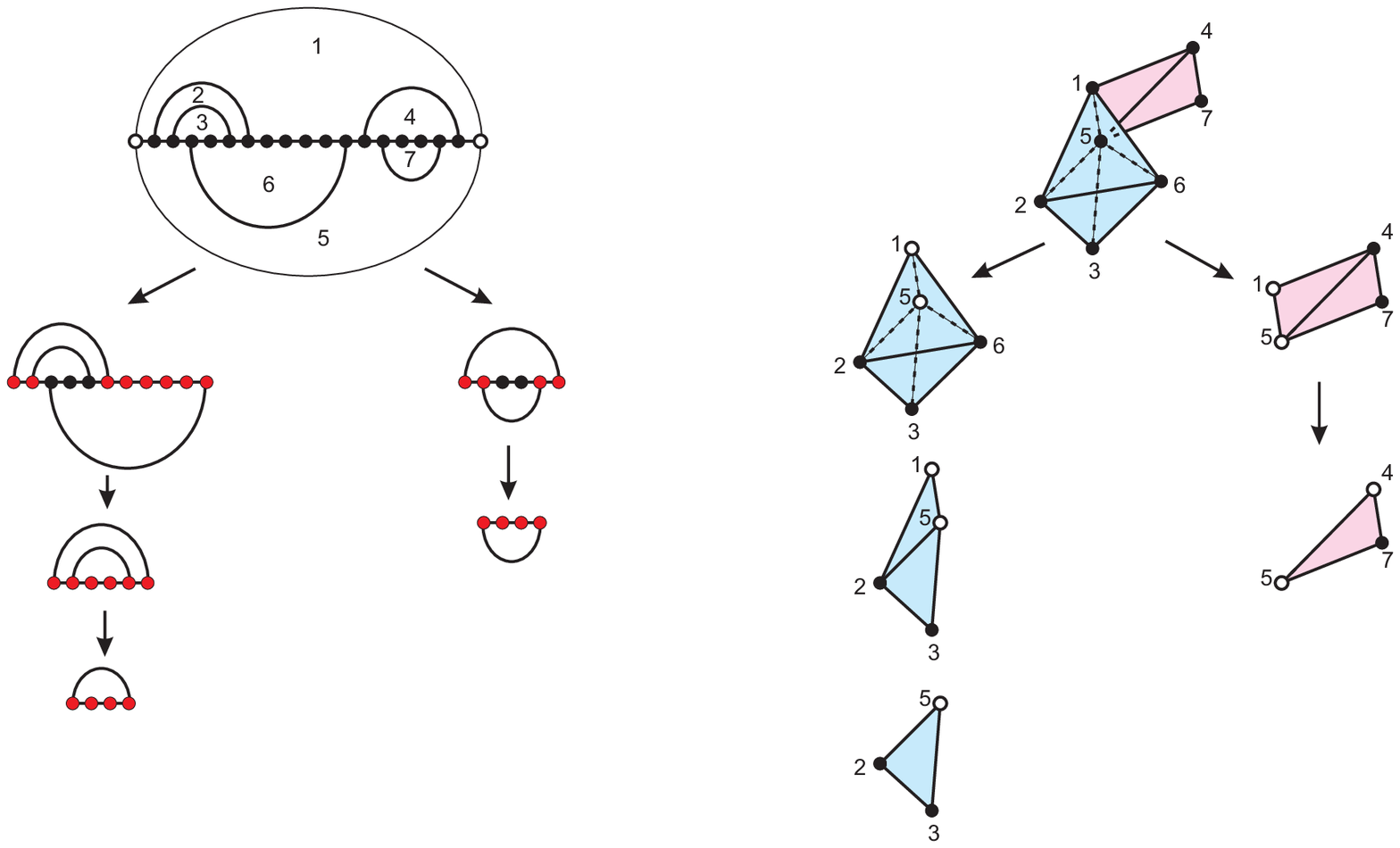}
\end{center}
\caption{\small A decomposition of a bistructure (LHS) and the evolution of its loop nerve (RHS). On the left the exposed vertices are marked in red. On the right the loops $1$, $2$, $3$, $5$ and $6$ forms 
a sphere. removing one loop is deleting one vertex of the loop nerve. The white vertices in the loop nerve are the boundary of the substructure. The sphere corresponding to the crossing component is resolved by removing the $S$-arc such that it becomes noncrossing. 
}
\label{F:example}
\end{figure}

We first give an overview of designing a decomposition of a bistructure based on the topological framework. \cite{Bura:19} shows that the space is classified by the rank of its second homology group $r_2(B)$, and comprised of ribbons glued to filled tetrahedra and spheres. 
Each sphere corresponds to a crossing component in the arc diagram, and their number is counted by the rank $r_2(B)$. We shall show that the \#-P hardness of the decomposition problem stems from the spheres, as the ribbons and tetrahedra are organized in a tree-like fashion. The global tree-like structure induces a tree decomposition naturally, while the sphere are resolved locally. To resolve the spheres 
we can map the problem to a known NP-problem such as, for instance, the traveling salesman problem (TSP). This allows us to solve the spheres via approximation algorithms of the TSP \cite{rego2011traveling}. We illustrate this idea in Figure~\ref{F:example}. 

\subsection*{Topological framework} \label{S:framework}

In the following we discuss and provide some details. 
\begin{definition}
Suppose $B(R, S)$ is a bistructure having $n$ loops $B = \{L_1,\dots,L_n\}$. We call $Y =\{ L_{i_0}, \ldots, L_{i_d}\}$ a $d$-simplex of $B$ 
if and only if $\bigcap_{k=0}^d L_{i_k} \neq \varnothing$. Let $K_d(B)$ be the set of all $d$-simplices of $B$. Then the nerve of $B$ is 
$$
K(B) = \dot{\bigcup}_{d=0}^{\infty} K_d(B)\subseteq 2^B. 
$$
\end{definition}

\begin{figure}[ht]
\begin{center}
\includegraphics[width=0.8\columnwidth]{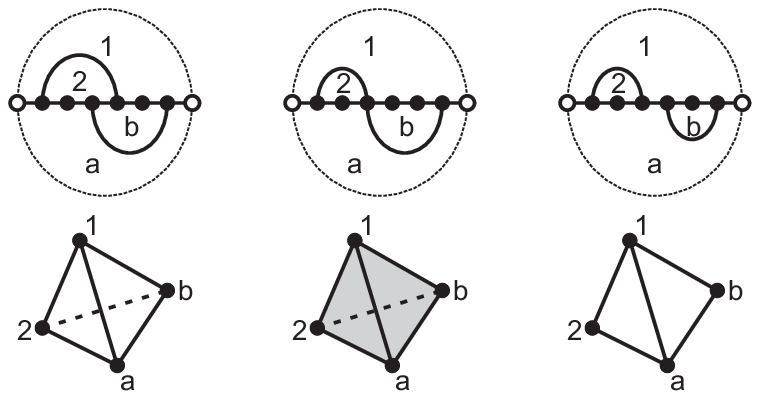}
\end{center}
\caption{\small Examples of topological realizations of the loop nerves for different bistructures: (A) an empty tetrahedron, (B) a filled tetrahedron, and (C) two filled triangles glued along a mutual edge. 
}
\label{F:nerve}
\end{figure}

The loop nerve $K(B)$ has the topological space $T(B)$ as its quotient space \cite{Massey:69}, see Figure~\ref{F:nerve}. 
The $0$-simplices correspond to hyper-edges in \cite{Hammer:19}. In the loop nerve the collection of $d$-simplices, captures the 
information of loop intersections not available in the hyper-graph model.

It is shown in \cite{Bura:19} that for a bistructure $B$, its loop nerve contains no $d$-simplex with $d>3$. 
Furthermore, there are only two nontrivial homology groups of $T(B)$, both free and abelian: $H_0(T(B)) \cong \mathbb{Z}$ and $H_2(T(B)) \cong \oplus_k^r \mathbb{Z}$. 
The rank of $H_2(T(B))$, denoted by $r_2(B)$, is thus the only parameter and so induces a natural classification. 
For an $R$-arc $(i,j)$ and an $S$-arc $(r,s)$ are crossing if $i<r<j<s$ holds. Next we discuss overlaps and crossing components.

\begin{figure}[ht]
\begin{center}
\includegraphics[width=0.9\columnwidth]{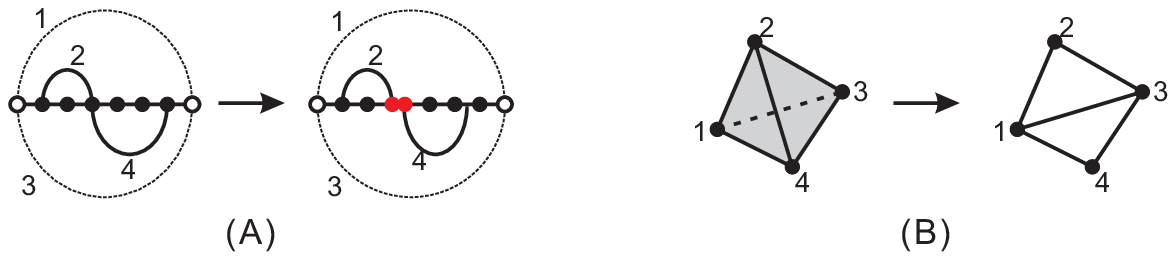}
\end{center}
\caption{\small Splitting an overlap without inducing crossing arcs (A). 
The split is tantamount to removing an edge of the corresponding filled tetrahedron as well as its interior, ending up with two triangles that are still glued along the opposite edge from the edge we removed.
}
\label{F:split}
\end{figure}

{\bf Overlaps:} 
an {\it overlap} is a degree four vertex in its arc diagram. An overlap corresponds to a $3$-simplex in $K_3(X)$ in the loop nerve. Assume $x$ is an 
overlap being the endpoint of the arcs $p_1 \in R$ and $p_2\in S$. We split $x$ into two adjacent vertices $x_1$ and $x_2$, where  $x_1$ carries the endpoint of $p_1$  and $x_2$ the endpoint of $p_2$ . This is done such that after the split $p_1$ does not cross $p_2$, see Figure~\ref{F:split} 

We investigate how the split affects the induced topological space. Let $x$ be an overlap. This vertex is contained in four loops and is the endpoint of two arcs $p_1$ and $p_2$. 
Let $L_1$ and $L_2$ be the loops in $R$ that contain $p_1$, where $p_1$ is the maximal arc of $L_2$. Furthermore let $L_3$ and $L_4$ be the loops in $S$ that contain $p_2$, where $p_2$ is the maximal arc of $L_4$. Clearly, $\cap_{i=1}^4 L_i = \{x\}$. It can be shown in the supplementary material (SM) that the action of splitting $x$ into $x_1$ and $x_2$ in a noncrossing fashion in the arc diagram is tantamount to removing an edge of the corresponding filled tetrahedron as well as its interior, ending up with two triangles that are still glued along the opposite edge from the edge we removed, see Figure~\ref{F:split}. 
This splitting does not change $r_2(B)$. Therefore, we can restrict ourselves to the non-overlap case.

{\bf Crossing components:} 
let $B(R,S)$ be the arc diagram of a bistructure with its arc set $\{p_1,\ldots, p_k\}$. We define the \emph{line graph} of $B$ to be $G=(B, E)$ where 
$E\ni e=(p_i, p_j)$ if and only if the arcs $p_i$ and $p_j$ cross. We call the set of arcs associated to a non-trivial connected component of this line graph, a \emph{crossing component} of $B$. 
By non-trivial we mean the vertex size of such a component is strictly larger than opne. We have 

\begin{lemma}\label{L:sphere}
Let $X^O$ be the substructure induced by a crossing component $O$, and let $\tilde{X^O}$ be its closure. Then
the induced topological space of $\tilde{X^O}$, $T(X^O)$, is homeomorphic to an empty sphere, and thus contributes $1$ to the rank of $H_2(B)$. 
\end{lemma}
 
We define the $*$-graph of the loop nerve to be the graph $\Delta(B)=(K_2(B),E)$ with edges given by 
$$
E\ni e=(\Delta_1,\Delta_2)\Leftrightarrow \Delta_1\cap \Delta_2\in K_1(B).
$$
Each vertex in the $*$-graph represents a filled triangle in $T(B)$, and there is an edge between two vertices if their respective triangles have nonempty intersection along an edge. Then we have:

\begin{lemma} \label{L:tree}
Let $X$ be a substructure without crossing arcs and overlaps, i.e., $H_2(B)=0$ and  $K_3(B)=\varnothing$. Then its $*$-graph $\Delta(B)$ is a tree.
\end{lemma}

\begin{figure}[ht]
\begin{center}
\includegraphics[width=0.8\columnwidth]{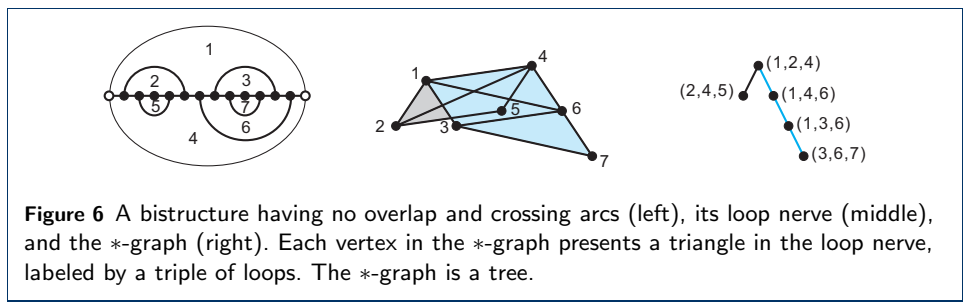}
\end{center}
\caption{\small A bistructure having no overlap and crossing arcs (left), its loop nerve (middle), and the $*$-graph (right). Each vertex in the $*$-graph presents a triangle in the loop nerve, labeled by 
a triple of loops. The $*$-graph is a tree. 
}
\label{F:pleat}
\end{figure}

We illustrate the $*$-graph of a  bistructure without overlaps and crossing arcs in Figure~\ref{F:pleat}. 
Note that, by Lemma~\ref{L:tree}, if $B$ has no crossings, the induced topological space $T(B)$ is a ``ribbon tree''. Namely, each ribbon is obtained by gluing a sequence of triangles along their edges such that each triangle has at most two edges glued to other triangles. These ribbons are then glued together along some of the edges of their constituent triangles such that no closed bands appear. 

Now we are in position to describe the structure of the topological space $T(B)$. If an irreducible substructure $X$ is induced by a crossing component, then the induced topological space is "sphere"-like. Otherwise if $X$ is noncrossing, the induced topological space is "ribbon tree"-like. $T(B)$ is a ribbon tree modulo edge contraction of spheres, see in the SM. Finally, we have the combinatorial interpretation 
of $r_2(B)$:

\begin{theorem}\label{T:class2}
Given a bistructure $B(R,S)$ with $r$ crossing components. then $r_2(B) = r$.  
\end{theorem}

The proofs of Lemma~\ref{L:sphere}, Lemma~\ref{L:tree}, and Theorem~\ref{T:class2} are presented in the SM.

\subsection*{Topological scheduling} \label{S:algo}

We next discuss how to design a decomposition based on the properties of the loop nerve. The global tree-like structure induces a tree decomposition naturally, while the sphere will be resolved locally. 
We first consider the case where $B$ contains no crossing arc. 
In this case, we extend the partial order $\prec$ for a bistructure by the following: for any two arcs $(i,j),(r,s)\in B$ we say $(i,j) \prec_B (r,s)$ if and only if $ i<r<s<j$. 
Then, we show in the SM that for an irreducible substructure $X\subseteq B$,  $X$ contains a unique maximal arc with respect to $\prec_B$. 

We decompose $X$ by removing the loop $L_{m}$, where $m$ is the maximal arc of $X$. The loop removal produces a set of irreducible substructure $X_1, \ldots, X_k$. 
Repeating this loop removal for any produced irreducible substructures gives a unique loop removal order $D_0(B)$. We can show in the SM that
\begin{lemma}\label{L:complexity0}
Let $B(R,S)$ be a bistructure without crossing arcs or overlaps. Let $D_0$  be the loop removal order discussed above. For any loop removal order $D\neq D_0$, we have $\kappa_{D_0}(B) \le \kappa_{D}(B)$, i.e., $D_0$ is a decomposition that minimizes $\kappa(B)$. 
\end{lemma}

A bistructure $B$ with overlaps can be mapped to a bistructures $B'$ without overlaps by the above splitting of overlapping vertices. The decomposition $D_0$ on $B'$ induces a natural 
decomposition $D$ on $B$ by the one-to-one correspondence between the $B$-arcs and the $B'$-arcs. We show in the SM that $D$ is optimal for $B$. 

We next discuss how to resolve spheres.  Recall that the \#-P hardness of the decomposition problem stems from the spheres. In this case, we consider mapping the problem to a known NP-problem as, for instance, the traveling salesman problem (TSP). To this end we remove a set of loops from $X$ with a minimum number of exposed vertices, such that $X$ has no crossing arcs. The remaining noncrossing substructure can be decomposed using the optimal algorithm presented before, see Figure~\ref{F:example}. 
This allows to solve the problem via approximation algorithms of the TSP \cite{rego2011traveling}. The approximation approach is the subject of future work and beyond the scope of this paper, for the analysis presented below, we employ a greedy approach to resolve the spheres. 

\section*{Results}

In this section we focus on using the RNA sequence sampler to study the sequence-structure relations of RNA riboswitch sequences. 
We display the RNA riboswitches that we investigate in Table~\ref{T:ribosw}. 

\begin{table}[h!]
\centering
\label{T:ribosw}
\tabcolsep=0.1 cm
\resizebox{\columnwidth}{!}{%
\begin{tabular}{|c|c|c|c|c|c|}
\hline
Abbr &  \text{ID} & Class & Organism & Reference  
\\ \hline
{\tt add} & add\_Adenine & Adenine & add - Vibrio vulnificus & \cite{add}  \\
{\tt xpt} & xpt\_Guanine & Guanine & xpt - Bacillus subtilis &  \cite{xptG}  \\
{\tt mgt} & mtgE\_Mg & Magnesium & mgtE - Bacillus subtilis &  \cite{Mgt_E} \\
{\tt lys} & lysC\_Lysine & Lysine & lysC - Bacillus subtilis & \cite{lysC}   \\
{\tt VEGFA} & VEGFA & Het. nuclear ribonucleoprotein L & VEGFA - Homo sapiens  &  \cite{VEGFA} \\
{\tt sam} & yitJ\_SAM & S-adenosylmethionine & yitJ - Bacillus subtilis &  \cite{Sam}  \\
\hline 
\end{tabular}
}
\caption{\small Data of riboswitch sequences that we investigate. }
\end{table}

\subsection*{Energy spectra}\label{S:spectrum}

Let us begin by introducing the spectrum over a partition function. Let $Q(X)$ be a partition function
of sequences compatible with $X$, where $X$ is a secondary or bistructure and $Q(X)=\sum_\sigma
e^{-\frac{\eta(\sigma, X)}{KT}}$. To simplify notation, we shall write $Q$ instead of $Q(X)$, if we do not
need to emphasize the context of the underlying structure $X$.
Naturally, $Q$ induces the discrete probability space $(\mathcal{Q}_4^n,\mathbb{P}_{Q})$, where 
$\mathbb{P}_{Q}(\sigma)=e^{-\frac{\eta(\sigma,X)}{KT}} / Q$.
We consider a real-valued random variable $f \colon (\mathcal{Q}_4^n,\mathbb{P}_{Q}) \longrightarrow
\mathbb{R}$ and refer to the induced measure $\mathbb{P}_{f}$ on $\mathbb{R}$,
$\mathbb{P}_{f}(r)=  \sum_{\{\sigma \mid f(\sigma)=r\}} \mathbb{P}_{Q} (\sigma)$, 
as the $f$-spectrum over $Q$. 

For practical purposes, an $f$-spectrum, $\mathbb{P}_f(r)$, cannot be computed directly, since
we have to consider all $\sigma \in \mathcal{Q}_4^n$ and potentially infinitely many $r\in \mathbb{R}$.
To approximate the $f$-spectrum we first discretize by means of a monotone
increasing sequence $(a_s)$, where $\Delta=a_s-a_{s-1}$, setting
$\mathbb{P}_f(a_s) = \sum_{a_{s-1}<r \le a_s} \mathbb{P}_f(r)$. 
We employ a Boltzmann sampler to generate sequences of the probability space
$(Q_4^n,\mathbb{P}_Q)$ and approximate $\mathbb{P}_{f}(a_s)$ by
$\mathbb{P}_{f}(a_s) \approx \frac{1}{m} | \{\sigma \mid a_{s-1} < f(\sigma) \le a_s\} |$, 
where $\sigma$ is a sequence sampled from the partition function $Q$ and $m$ denotes the sample
size. Here we set $m=10^4$.  

We proceed by introducing some particular choices for the pair $(f,Q)$, which we shall
denote by $f_Q$:
$$
f^R_{Q}(\sigma)=\eta(\sigma,R) \quad \quad f^S_Q(\sigma)= \eta(\sigma, S). 
$$ 
We call $\mathbb{P}_{f_Q^R}(r)$ the {\it $R$-spectrum of $Q$} and $\mathbb{P}_{f_Q^S}(r)$ the
{\it $S$-spectrum of $Q$}. 

We are now in position to study the $R$- and $S$-spectra of $Q$-Boltzmann sampled, bicompatible
sequences of specific structure pairs, or equivalently the $R$- and $S$-spectra of sequences
compatible to bistructures.
It will be interesting to compare these with the spectra of the compatible sequences of each
respective secondary structures, $R$ and $S$ and to provide a comparative analysis of the $R$-
and $S$-spectra of native structures with that of random structure pairs.

\begin{figure}
\begin{tabular}{ccc}
\includegraphics[width=0.4\columnwidth]{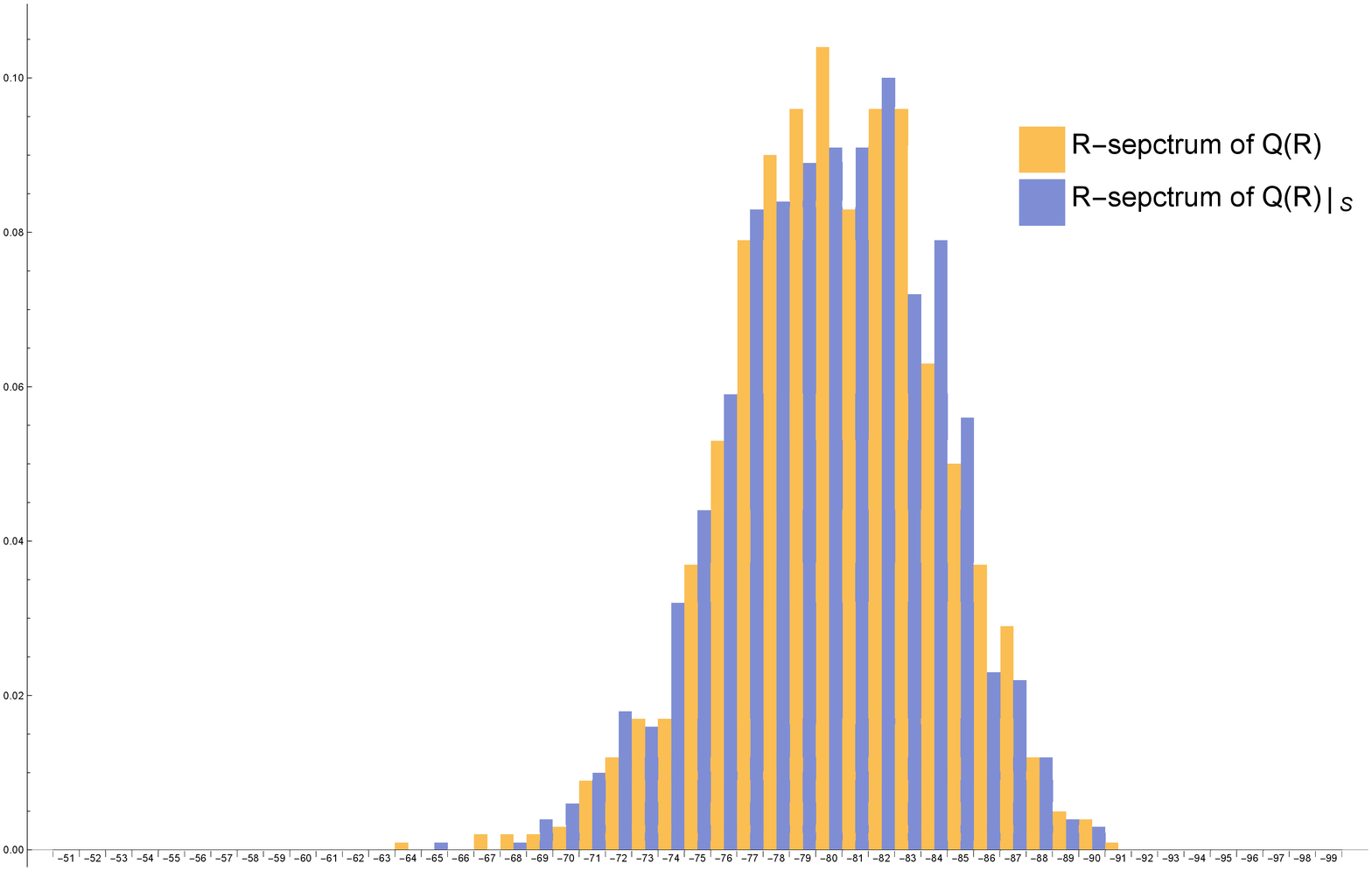} &
\includegraphics[width=0.4\columnwidth]{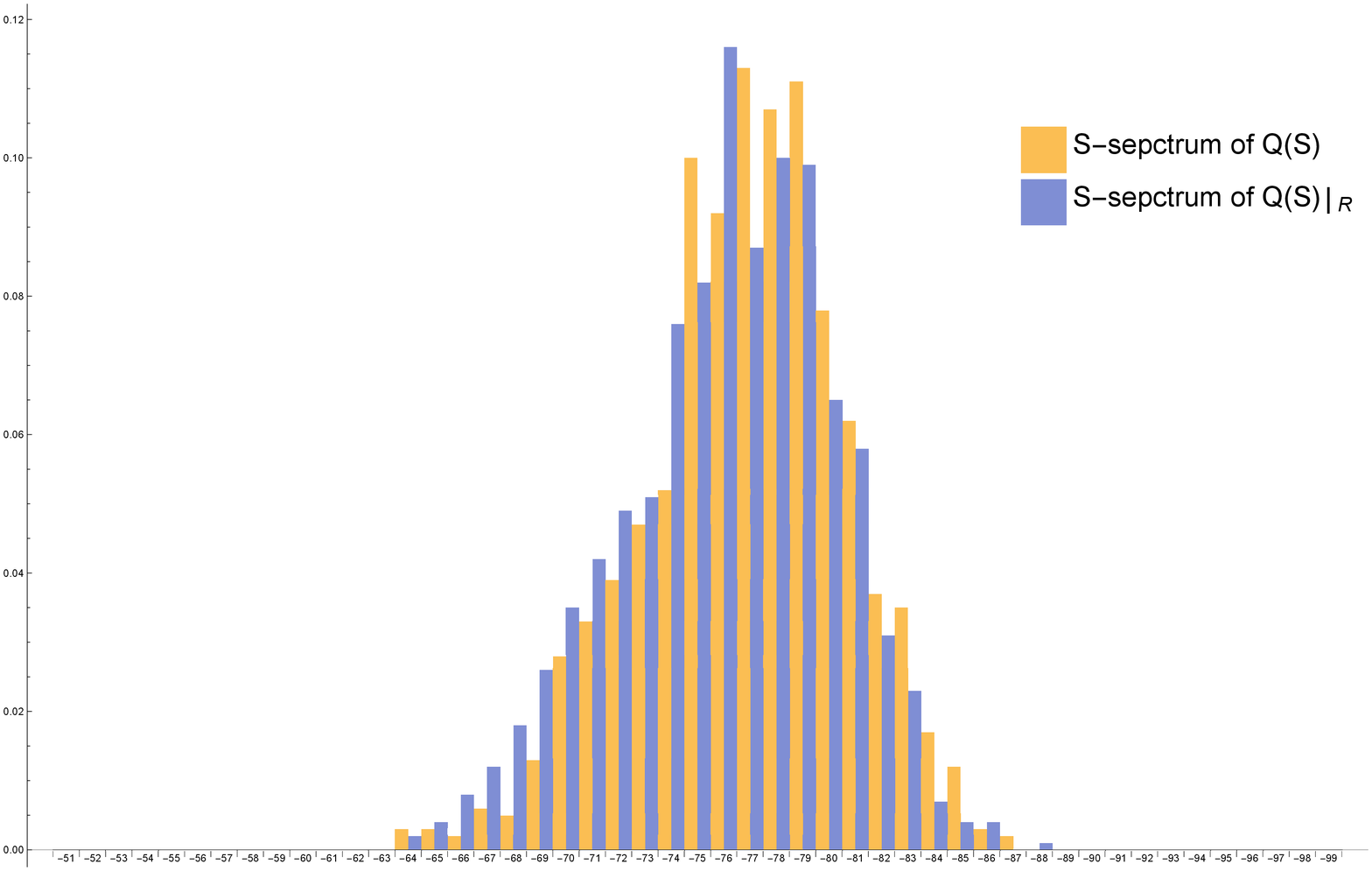} \\
$R$-spectrum of {\tt add} & $S$-spectrum of {\tt add} \\
\includegraphics[width=0.4\columnwidth]{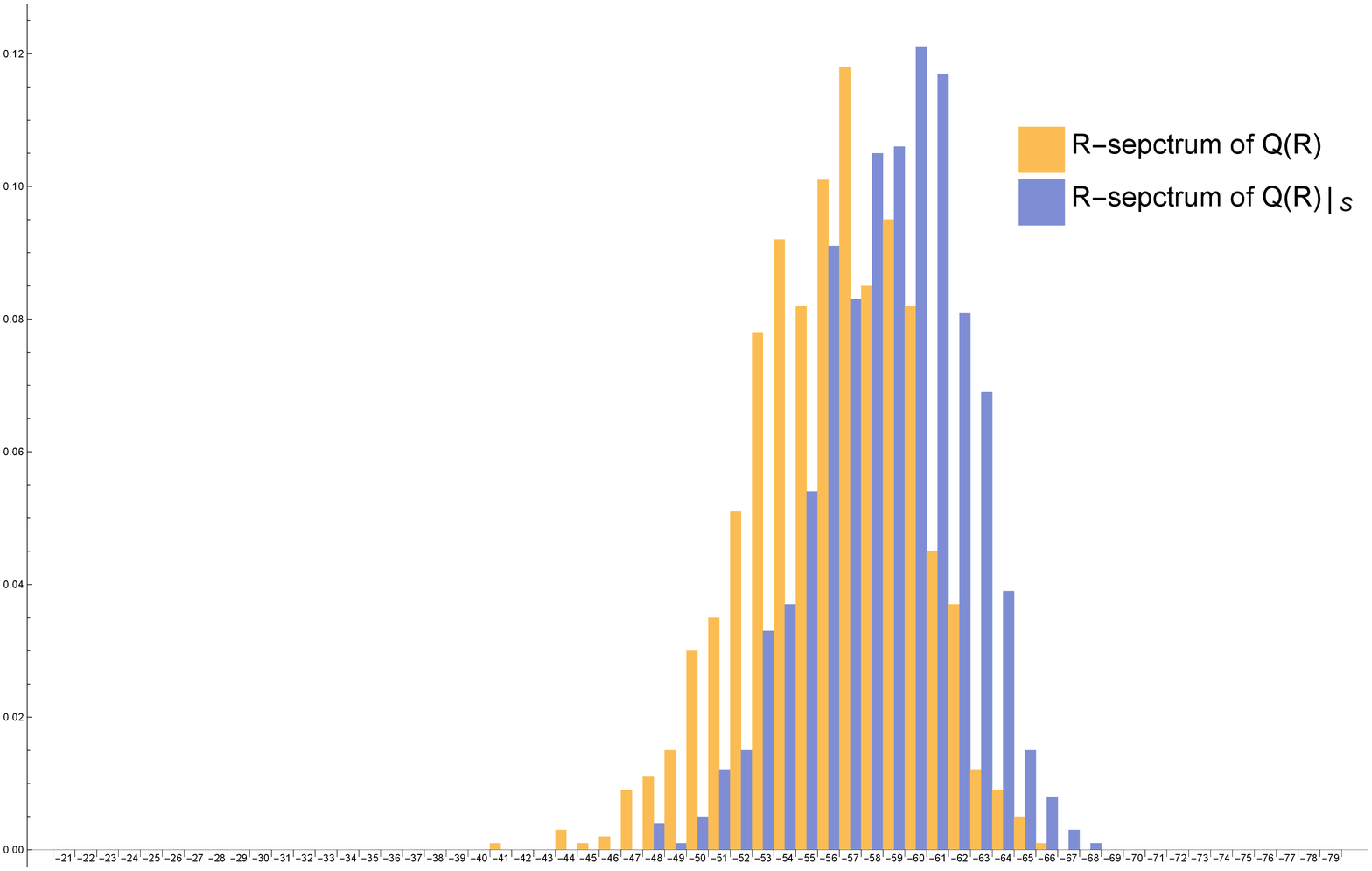} &
\includegraphics[width=0.4\columnwidth]{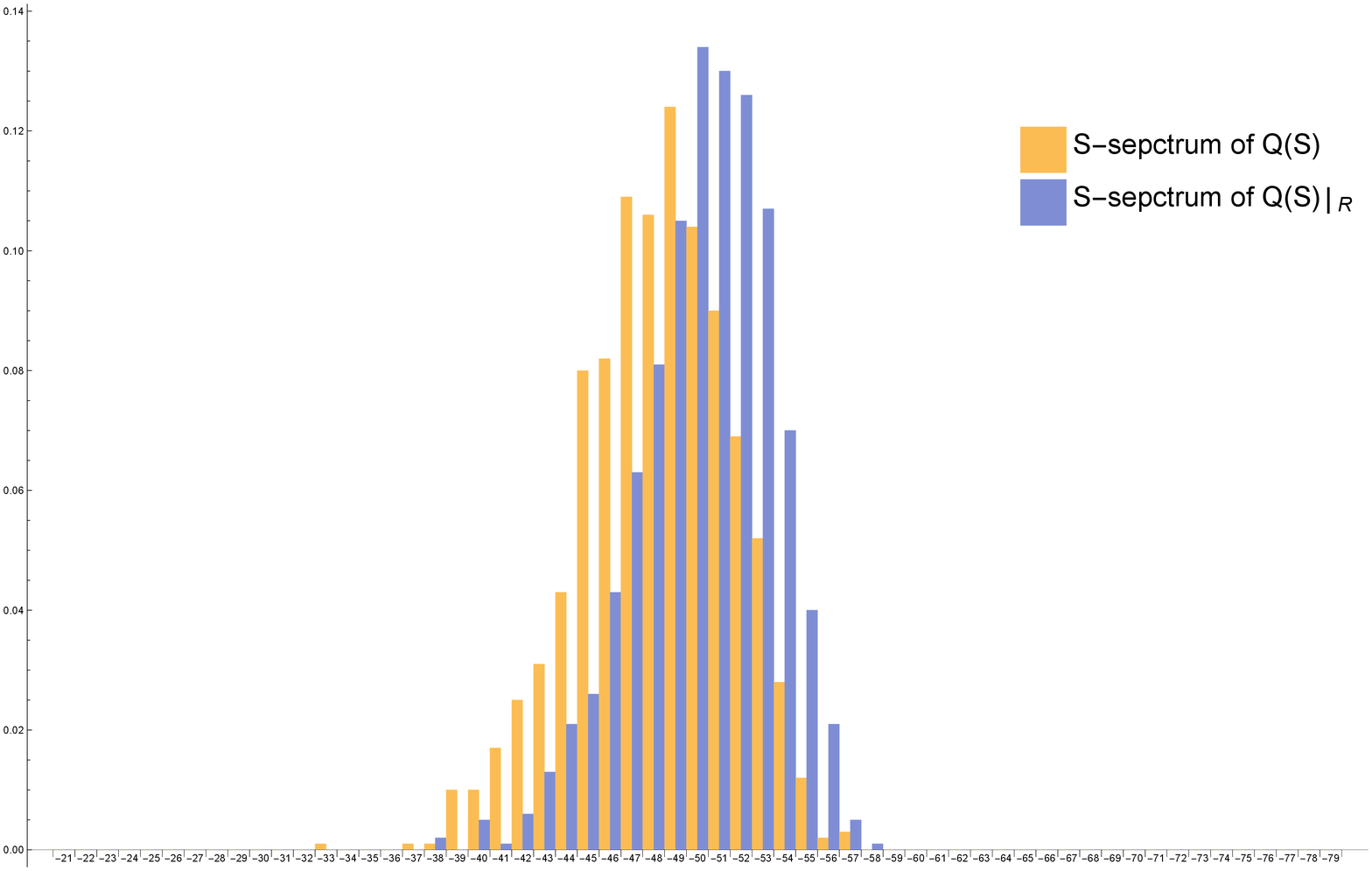} \\
$R$-spectrum of {\tt rand1} & $S$-spectrum of {\tt rand1}
\end{tabular}
\caption{\small The $R$- and $S$ spectra of the riboswitch structure pair, {\tt add} and {\tt rand1}.
upper left: $f_{Q(R)}^R$ versus $f_{Q(R)|_S}^R$ for {\tt add}.  
upper right: $f_{Q(S)}^S$ versus $f_{Q(S)|_R}^S$ for {\tt add}.  
lower left:  $f_{Q(R)}^R$ versus $f_{Q(R)|_S}^R$ for {\tt rand1}.  
lower right: $f_{Q(S)}^S$ versus $f_{Q(S)|_R}^S$ for {\tt rand1}.  
}
\label{F:add_ind}
\end{figure}

Let $R,S$ be two secondary structures, to begin the comparative analysis of compatible and
bicompatible sequences we compare $Q(R)$ and $Q(S)$, given by 
$$
Q(R) = \sum_{\sigma\in \mathcal{Q}_4^n} e^\frac{{-\eta(\sigma,R) }}{KT},  \quad \quad 
Q(S) = \sum_{\sigma\in \mathcal{Q}_4^n} e^\frac{{-\eta(\sigma,S) }}{KT}, 
$$
with the partition functions
$$
Q(R)|_S = \sum_{\sigma\in \mathbb{C}_n(R, S)} e^\frac{{-\eta(\sigma,R) }}{KT},  \quad 
Q(S)|_R = \sum_{\sigma\in \mathbb{C}_n(R,S)} e^\frac{{-\eta(\sigma,S) }}{KT}. 
$$

$Q(R)$ and $Q(S)$ are recursively computed and their Boltzmann samplers are introduced in
\cite{Clote:16, Barrett:17}. On an abstract level, $Q(R)|_S$ and $Q(S)|_R$ were \emph{a priori} available
by means of rejection Boltzmann samplers for $Q(R)$ and $Q(S)$.
However, sampling such sequences is impractical as the probability of
randomly encountering a bicompatible sequence is too low.
As a first application, our framework developed in Section~\ref{S:method} we observe that $Q|_S(R) $ and
$Q|_R(S)$ can be computed by replacing the energy function 
$\eta(\sigma, B(R,S)$ by $\eta(\sigma, R)$ and $\eta(\sigma, S)$, respectively.

Comparing the $R$- and $S$-spectra of $Q(R)$ with $Q(R)|_S$ and $Q(S)$ with $Q(S)|_R$, respectively,
allows us to draw conclusions about the difficulty of the process of modifying a $R$-compatible
sequence into an $R,S$-bicompatible sequence, while maintaining the energy with respect to $R$ and the
analogue statement for $S$.

In Figure.~(\ref{F:add_ind}) we compare $f_{Q(R)}^R$ with $f_{Q(R)|_S}^R$,
i.e.~the $R$-energy spectra over $Q(R)$ and $Q(R)|_S$ (LHS) and $f_{Q(S)}^S$ with
$f_{Q(S)|_R}^S$, i.e.~the $S$-energy spectra over $Q(S)$ and $Q(S)|_R$ (RHS). We remark that these are
pairwise comparisons of measures over distinctively different, nested probability spaces.

We show that the $R$-energy spectra over $Q(R)$ and $Q(R)|_S$ and $S$-energy spectra over $Q(S)$ and $Q(S)|_R$
pairwise coincide. This means, that modifying a $R$-compatible sequence into a $R,S$-bicompatible sequence
can be done without affecting the free energy with respect to $R$ and vice versa for $S$. Moreover, this finding
holds for all native, as well as random structure pairs we analyzed. The results for the other riboswitch as well 
as random structures pairs are shown in the SM.

\begin{figure}
\begin{tabular}{ccc}
\includegraphics[width=0.4\columnwidth]{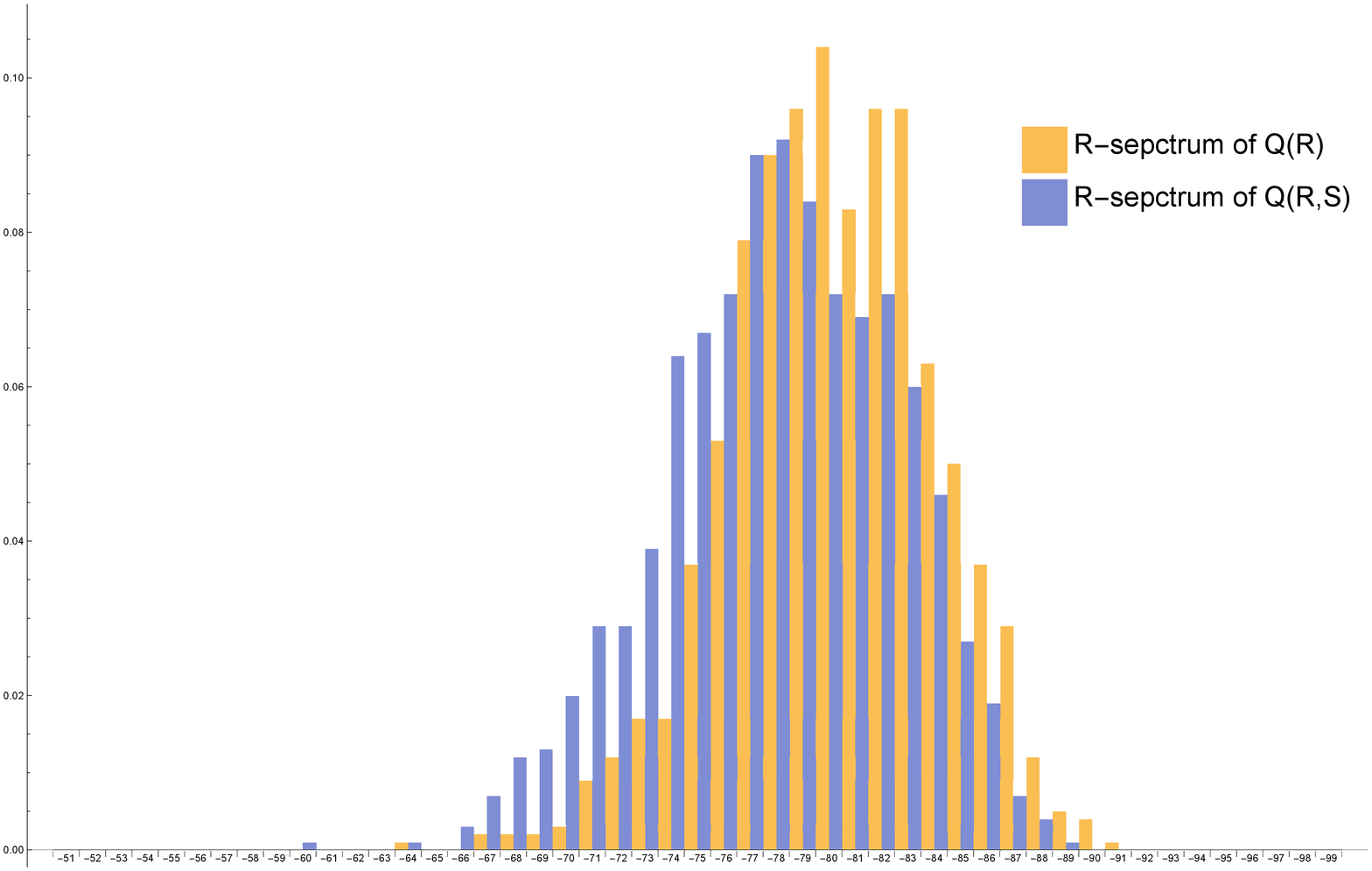} &
\includegraphics[width=0.4\columnwidth]{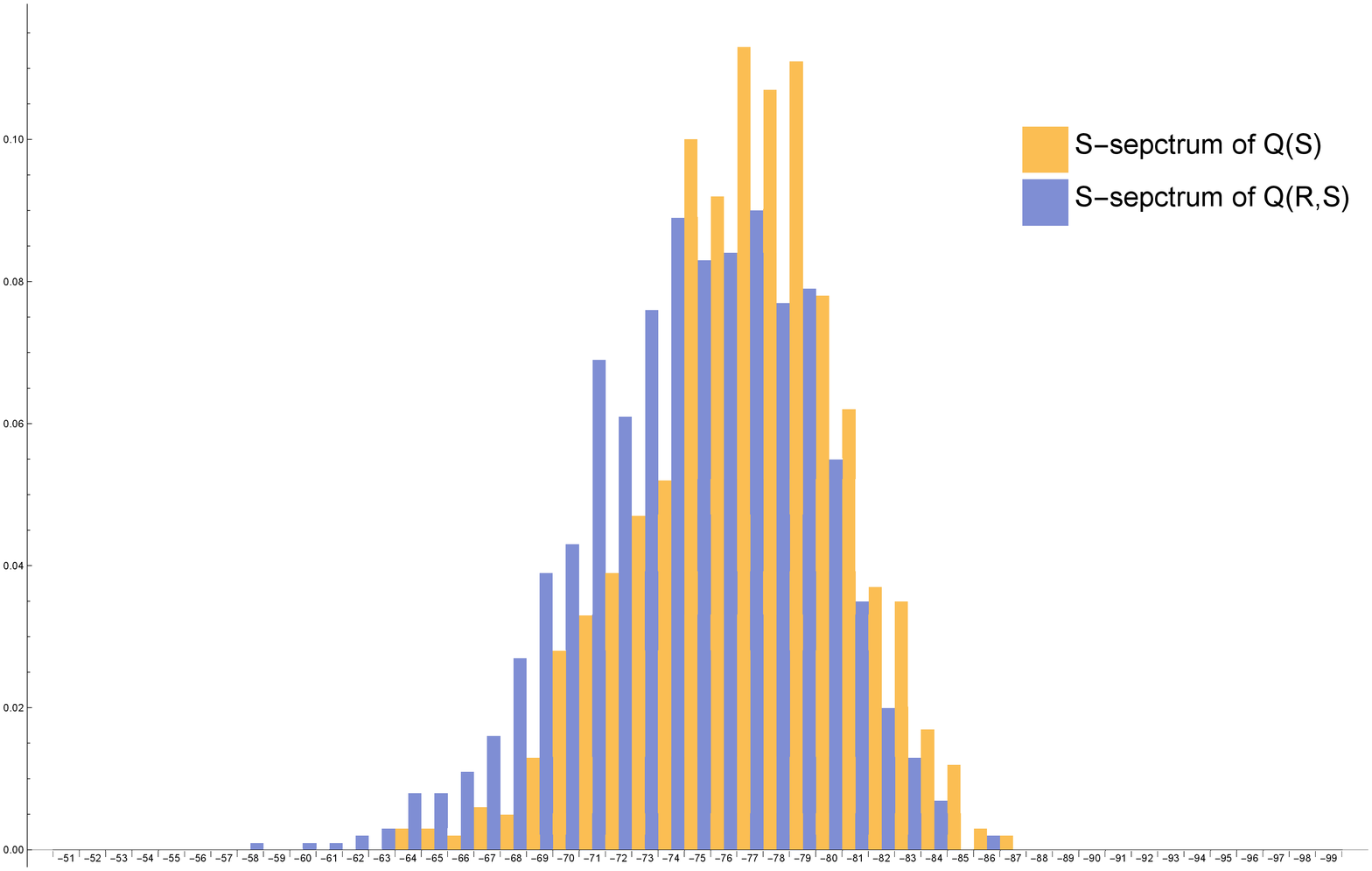} \\
R-spectrum of {\tt add} & S-spectrum of {\tt add}  \\
\includegraphics[width=0.4\columnwidth]{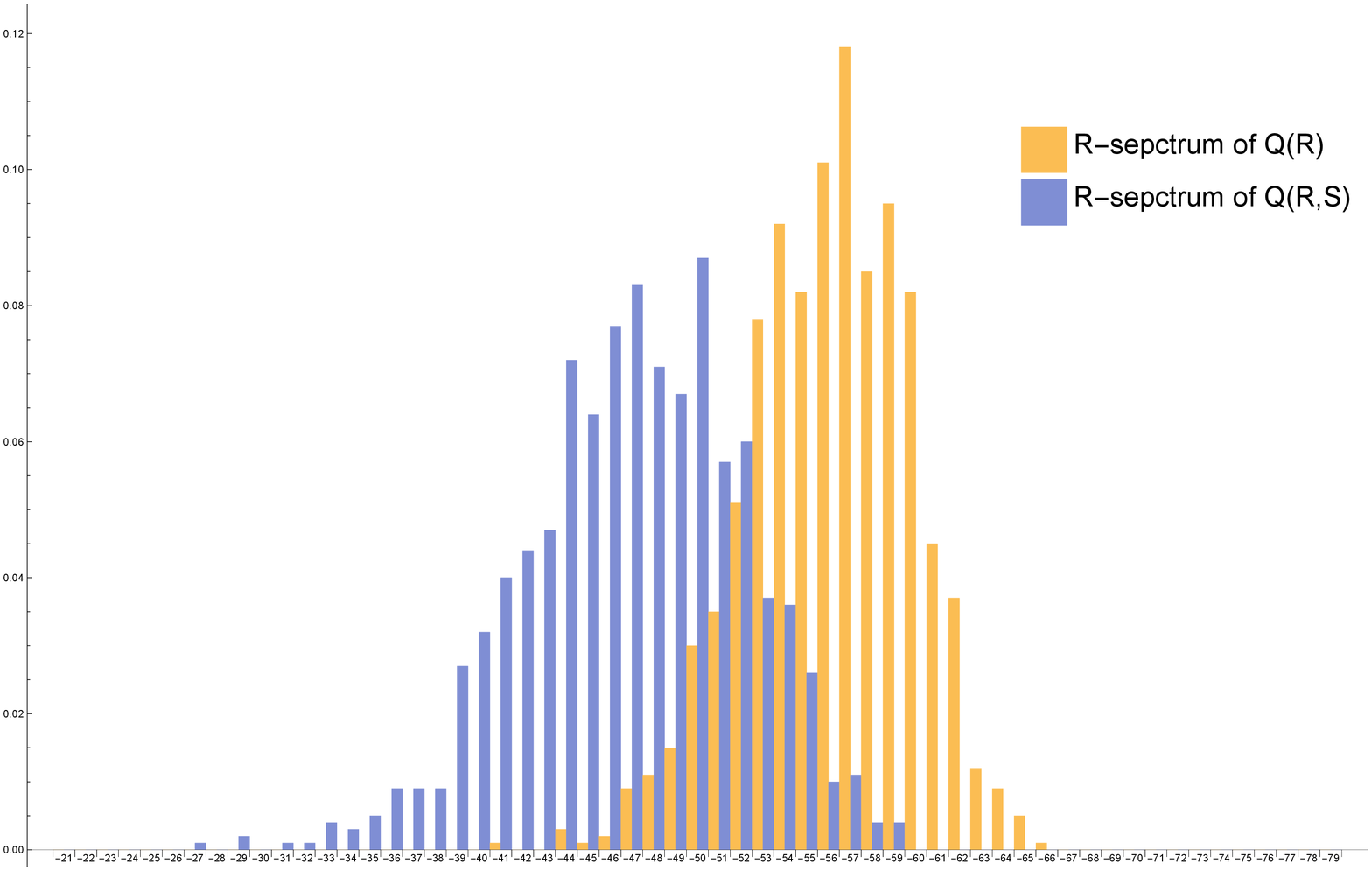} &
\includegraphics[width=0.4\columnwidth]{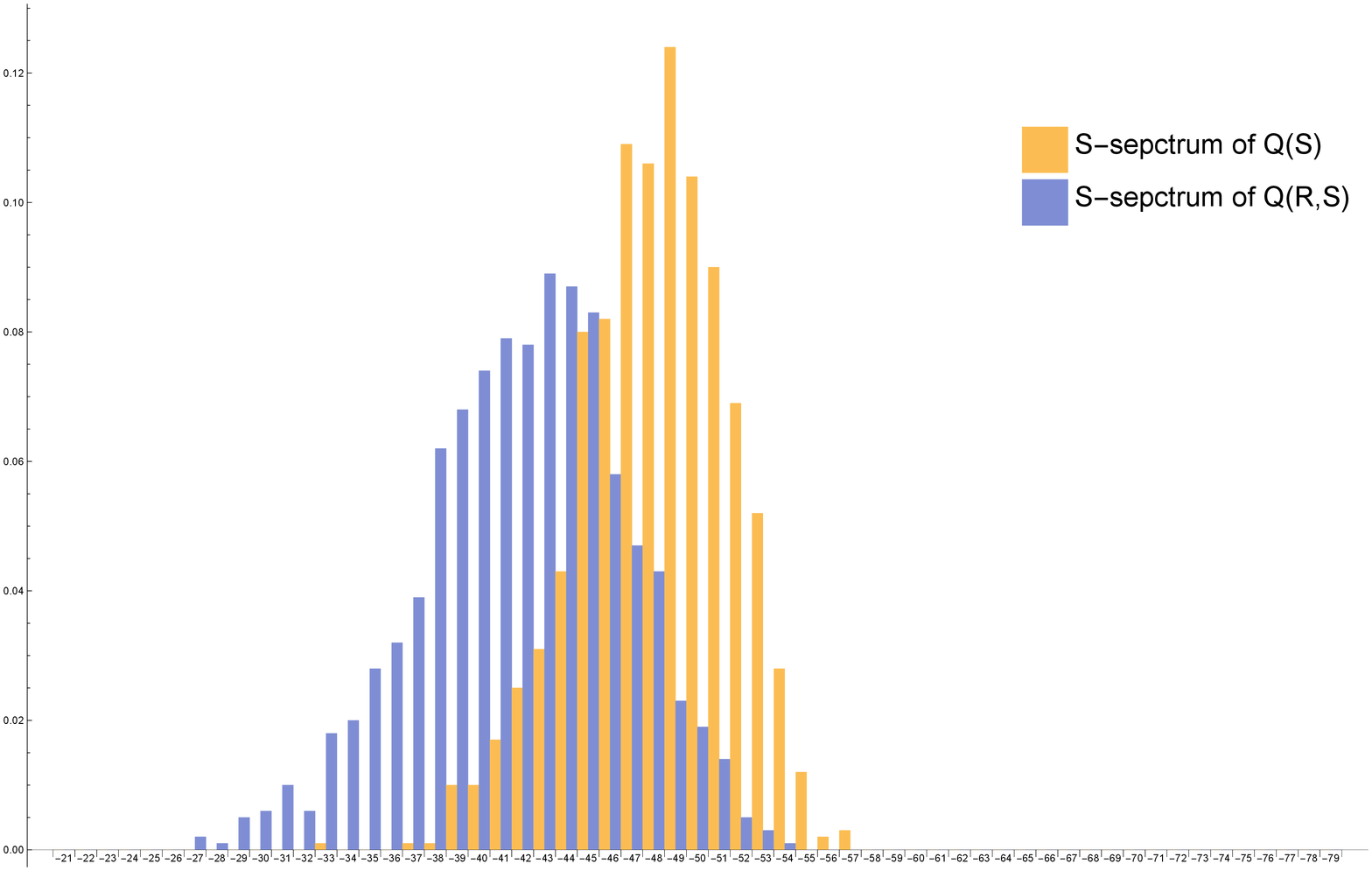} \\
R-spectrum of {\tt rand1} & S-spectrum of {\tt rand1} 
\end{tabular}
\caption{\small The $R$- and $S$ spectra of the riboswitch structure pair, {\tt add} and {\tt rand1}.
upper left: $f_{Q(R)}^R$ versus $f_{Q(R,S)}^R$ for {\tt add}.  
upper right: $f_{Q(S)}^S$ versus $f_{Q(R,S)}^S$ for {\tt add}.  
lower left:  $f_{Q(R)}^R$ versus $f_{Q(R,S)}^R$ for {\tt rand1}.  
lower right: $f_{Q(S)}^S$ versus $f_{Q(R,S)}^S$ for {\tt rand1}.  
}
\label{F:add}
\end{figure}

The next step is to relate bicompatible sequences, Boltzmann sampled from $Q(R,S)$ to those Boltzmann sampled
from $Q(R)$. In the context of the $Q(R)|_S$ versus $Q(R)$ analysis, we now factor in the energy with respect to
$S$. While $R$-energy levels can be maintained while satisfying the $S$ base pairing conditions, it turns out to
be much more intricate to derive bicompatible sequences that are well suited for both $R$ and $S$ at the
same time. Thus we consider
$$
Q(R,S) = \sum_{\sigma\in \mathbb{C}_n(R, S)} e^\frac{{-1/2 \cdot (\eta(\sigma,R) + \eta(\sigma,S))}}{KT}
$$
and compare the energy spectra $f_{Q(R,S)}^R$ and $f_{Q(R)}^R$ as well as $f_{Q(R,S)}^S$ and $f_{Q(S)}^S$
for a variety of riboswitch sequences and their respective native structure pairs, see Table~\ref{T:ribosw}.
We display the energy spectrum of the riboswitch {\tt add} in Figure~\ref{F:add} (upper). The energy spectra and
detailed analysis of the other riboswitch sequences is presented in the SM. 
In order to put our results into context, we present an analysis of the spectra
$f_{Q(R,S)}^R$ and $f_{Q(R)}^R$ as well as $f_{Q(R,S)}^S$ and $f_{Q(S)}^S$ for random structure pairs.
Here, as described above, random means we consider a pair of mfe-secondary structures of two randomly selected
sequences of length $100$. We show a representative result for a random structure pair in Figure~\ref{F:add} (lower).
Our observations are remarkably robust: the energy spectra of random structure pairs are literally
identical and we provide a detailed analysis of additional spectra in the SM.

Figure~\ref{F:add} (upper left) shows that the $R$-spectrum, $f_{Q(R,S)}^R$, is practically identical to the
$R$-spectrum, $f_{Q(R)}^R$. The $S$-spectra behave completely analogous, there is no significant
difference between the $S$-spectrum $f_{Q(R,S)}^S$ and $f_{Q(S)}^S$, see Figure~\ref{F:add} (upper right).
In the SM we provide additional data and the spectra of the riboswitch sequences listed in
Table~\ref{T:ribosw}. The phenomenon holds robustly for all riboswitch sequences we analyzed.

For random structure pairs, however, the picture changes: the $R$-spectrum, $f_{Q(R,S)}^R$, is shifted
distinctively to the left of the $R$-spectrum, $f_{Q(R)}^R$, see Figure~\ref{F:add} (lower left). The same
holds for the $S$-spectra: $f_{Q(R,S)}^S$ is shifted distinctively to the left of the $S$-spectrum
$f_{Q(S)}^S$, see Figure~\ref{F:add} (lower right). More data for random structure pairs are presented in the SM.

\subsection*{Ranking} \label{S:ranking}
\begin{figure}[ht]
\begin{tabular}{ccc}
\includegraphics[width=0.4\columnwidth]{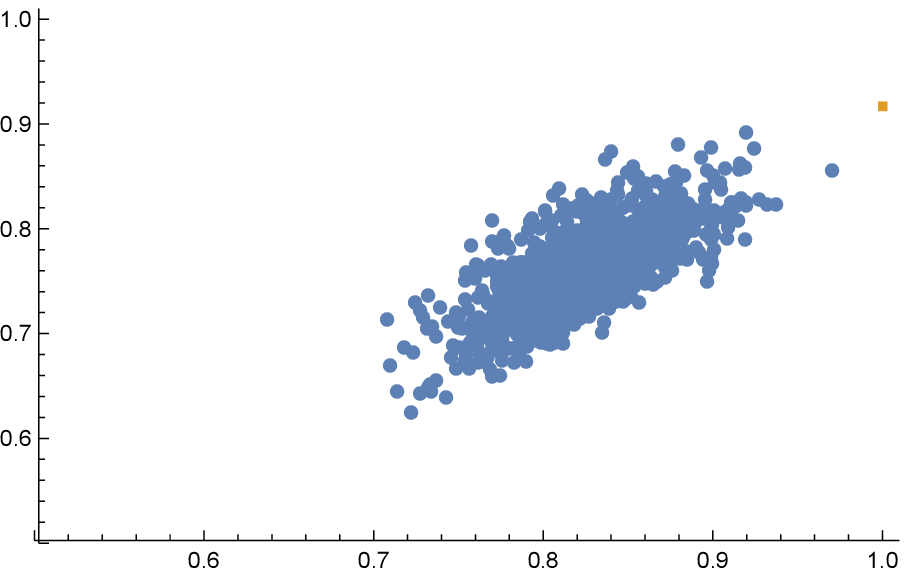} &
\includegraphics[width=0.4\columnwidth]{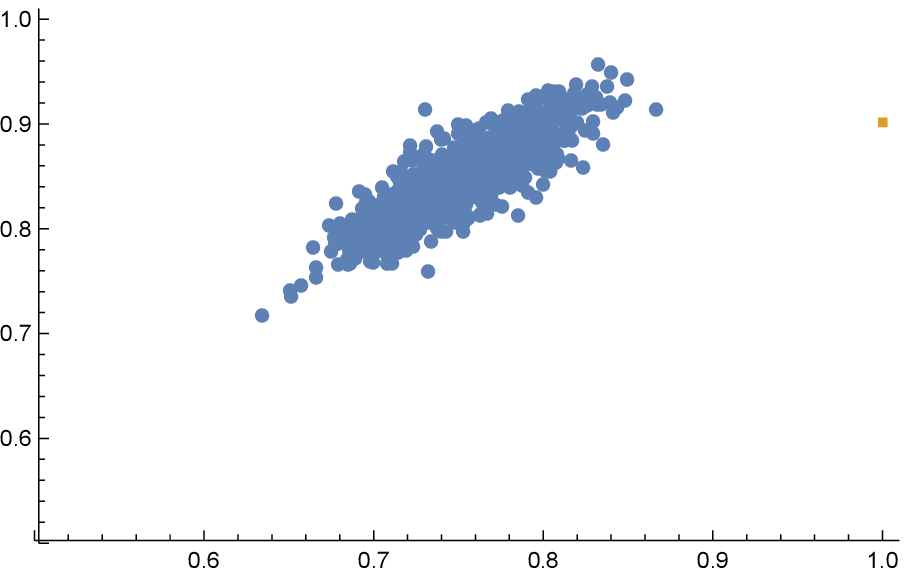} \\
{\tt mgt} & {\tt lys}
\end{tabular}
\caption{\small
$(r_R, r_S)$ for the riboswitches {\tt mgt} (left) and {\tt lys} (right): Boltzmann sampled sequences versus
  native sequences. For each riboswitch we Boltzmann sample $10^3$ sequences and compute $(r_R, r_S)$ for the
  sampled sequences (blue). We contrast this with $(r_R, r_S)$ of for the respective, native riboswitch
  sequence (red). 
 }
\label{F:ranking}
\end{figure}

Let $\sigma\in (\mathcal{Q}_4^n,\mathbb{P}_{Q(R,S)})$, i.e.~$\sigma$ is Boltzmann sampled sequence via $Q(R,S)$.
In this section, we compare the energies, $\eta(\sigma, R)$ and $\eta(\sigma, S)$ to $\eta(\sigma, M(\sigma))$,
where $M(\sigma)$ denotes the mfe-structure of $\sigma$. We consider the ratios
$$
r_R = \frac{\eta(\sigma, R)}{\eta(\sigma, M(\sigma))}, \quad \quad
r_S = \frac{\eta(\sigma, S)}{\eta(\sigma, M(\sigma))},
$$
which reflect the similarity between the energies of $R$ and $S$ with the minimum free energy.
We display $(r_R, r_S)$ for the riboswitches {\tt mht} and {\tt lys} in Figure~\ref{F:ranking}. 

Figure~\ref{F:ranking} displays the ratios $(r_R, r_S)$ for the riboswitch {\tt mgt} (LHS) and {\tt lys} (RHS).
The figure is obtained based on the Boltzmann sampling of $10^3$ sequences from $(Q(R,S))$ (blue) and displays
in addition the ratios of the native sequences of {\tt mgt} and {\tt lys} (red).

We find for {\tt mgt} that all ratios satisfy $r_R > 70\%$ and $r_S > 62\%$ and furthermore the respective
(coordinatewise) means are $(85\%, 74\%)$. For {\tt lys} we have $r_R > 64\%$ and $r_S > 72\%$ with a mean
of $(78\%, 85\%)$. Accordingly, $R$ and $S$ are suboptimal structures in the Boltzmann sampled sequences. 
Furthermore, we find that the ratio of ratios, $r_R/r_S$ is almost constant within the set of sampled sequences.
For both, {\tt mgt} and {\tt lys} alike, $(r_R, r_S)$ of the native sequence is distinctively higher than the
ratio pairs obtained from the sampled sequences. This indicates that the native sequence exhibits an
evolved thermodynamic stability with respect to the pair of structures $(R,S)$.

\subsection*{Density}\label{S:density}

In Section~\ref{S:spectrum}, we discuss the energy-spectrum over a partition function, $Q$ as an
induced measure of a random variable, $f$. By construction we normalize, when working with the probability
measure $\mathbb{P}_{Q}(\sigma)$, the value of $Q$. As a result, the absolute values of the different
partition functions, for instance, when comparing $Q(R)$ and $Q(R)|_S$ is not a factor.

Comparing a plethora of riboswitches, as well as sequences of various random structure pairs,
we end up with relating the partition functions of sequences over an entire spectrum of lengths.
The free energy of any sequence is however the sum of loop energies, and each loop has a unique
maximal arc. In \cite{Jin:08b} it is shown that the number of arcs in random structures satisfies a central limit
theorem, whence its mean scales linearly with $n$. This implies that the number of loops grows linearly
with $n$, which in turn suggests that the free energy of a sequence grows linearly with $n$.

\begin{figure}[ht]
\begin{center}
\includegraphics[width=0.8\columnwidth]{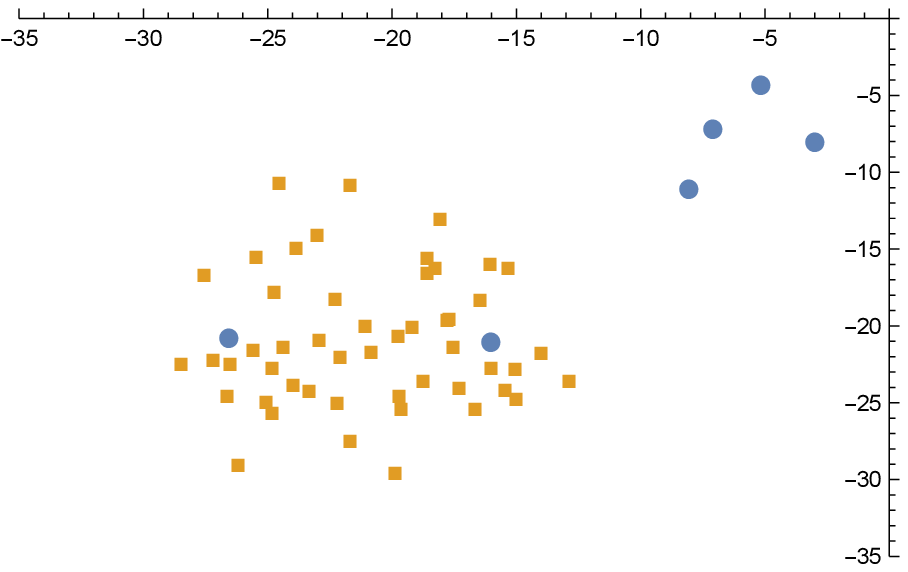}
\end{center}
\caption{\small The energy weighed space of bicompatible sequences. $(w_R,w_S)$ of the six riboswitch
  structure pairs, presented in Table~\ref{T:ribosw} (blue). Furthermore: $(w_R,w_S)$ of $10^2$ random
  structure pairs of length $100$ (red). 
}
\label{F:PFratio}
\end{figure}

Accordingly, we consider the scaled partition function
$$
\tilde{Q}(R)=\sum_{\sigma}e^{\frac{1}{n}\frac{\eta(\sigma,R)}{KT}} \quad \quad
\tilde{Q}(R)|_S=\sum_{\sigma \in \mathbb{C}_n(S)} e^{\frac{1}{n}\frac{\eta(\sigma,R)}{KT}}
$$
and set
$$
w_R = \log\left(\frac{\tilde{Q}(R)|_S}{\tilde{Q}(R)}\right), \quad \quad w_S = \log\left(\frac{\tilde{Q}(S)|_R}{\tilde{Q}(S)}\right). 
$$
We compute in Table~\ref{T:ribosw}, $(w_R,w_S)$ for the riboswitch structures pairs, and augment the analysis
by inspecting $50$ random structure pairs. We display in Figure~\ref{F:PFratio} the pairs $(w_R,w_S)$. 

Figure~\ref{F:PFratio} shows that the pairs $(w_R,w_S)$ of riboswitch sequences are distinctively different
from random structure pairs and appear in the upper right corner, while the $(w_R,w_S)$ of random structure
pairs are shifted towards the lower left corner.
The closeness of ratio pairs displayed in Figure~\ref{F:PFratio} to the upper right corner represents how
likely a sequence, Boltzmann sampled from $\tilde{Q}(R)$, is contained in $\tilde{Q}(R)|_S$. This reflects how dense the
bicompatible sequences sampled from $\tilde{Q}(R)$ are within the compatible sequence Boltzmann sampled from $\tilde{Q}(R)$.
Figure~\ref{F:PFratio} shows that this density is significantly higher for native riboswitch sequences, compared
to sequences Boltzmann sampled from $\tilde{Q}(R)|_S$ for random structure pairs $(R,S)$.

\section{Discussion}

The time complexity of computing the partition function for a given pair of secondary structures is determined by the decomposition of the bistructure and the computation of 
the partition function. The former is \#P-hard, and the latter is of polynomial time complexity using a DP-routine. Accordingly, the problem is FPT. 
\cite{Hammer:19} models the former using hyper-graphs and focuses on the latter, presenting an algorithm that computes the partition function on a simplified loop-based energy model.
Although in general a hyper-graph can be decomposed into a tree-like structure having tree-width $k$, the parameter $k$ is not understood in a systematic way. 

We notice that interactions of loops in two secondary structures are not arbitrary, namely, they have to satisfy specific constraints. Hyper-graph models lack of the capability 
to adequately capture this additional information. However, interpreting the loop-intersections as a simplicial complex, \cite{Bura:19} unlocks a homological framework for the analysis of bistructures.
The topology captures higher order information by higher dimensional simplices. Theorem~\ref{T:class2} provides a natural classification of bistructures via the rank of their 
second homology group, $r_2(B)$. It further describes the topological space of a given bistructure, namely, a ribbon tree modulo the contraction of spheres. 
In Lemma~\ref{L:complexity0} we construct in case of $r_2(B)=0$, i.e., for a ribbon tree, the optimal loop removal schedule.

Each sphere is given a concrete interpretation, 
namely, as a crossing component, and determined by $r_2(B)$. This understanding of the topological space provides insight into the design of decompositions of the bistructure. 
The tree-like structure is naturally a tree decomposition, while the complexity of the problem originates from the spheres. Though resolving the sphere is \#P-hard, it is possible 
to map it to a known NP-problem like, for instance the traveling salesman problem (TSP). Via such a mapping, efficient approximation algorithms \cite{rego2011traveling}
can be levied. 

The filtration of the sequence space by mapping sequences into their mfe-secondary structure has been a powerful
tool for the analysis of evolutionary optimization \cite{Schuster:94}.
This search exhibits extended periods of phenotypic
neutrality separated by transition events, during which structural change manifests
\cite{Fontana:98}. Bicompatible sequences
facilitate these transitions, which depend heavily on the particular choice of the two structures.
Our framework allows us to quantify the accessibility of the two structures by means of the density ratio
$(w_R, w_S)$, see Figure~\ref{F:PFratio}. The result shows that this ratio is distinctively
higher for the native structure pairs of riboswitches compared to random structure pairs. As a result
transitions between native structure pairs are much easier than for random pairs. This motivates to
identify what properties of the native structure pairs lead to the high density of their bicompatible
sequences.   

Over two decades multistable sequences were analyzed \cite{Flamm:01}. These are multistable to alternative
conformations, performing different functionality by switching between their alternative structures in
gene regulation and as such these sequences represent beacons in evolution.
\cite{Flamm:01} studies sequences that are multistable with respect to two structures, both being suboptimal by mapping
the problem into a combinatorial optimization problem. The latter is then solved via an adaptive walk,
initiated at a random sequence. The Boltzmann sampling of the starting sequences for the adaptive walks
of \cite{Flamm:01} will likely have the same speedup effect, as the Boltzmann sampling of compatible
sequences in the context of inverse folding \cite{Busch:06}.

On the level of phenotypes, our results on energy spectra and density of a bistructure can clearly distinguish between riboswitch or native and random structure pairs reliably, see 
Figure~\ref{F:add} and Figure~\ref{F:PFratio}.
While we are not yet in position to formulate criteria for designing such structure pairs, we can recognize them. 
As for genotypes, it is not easy to identify whether a sequence is a candidate for a riboswitch--even
if the native structure pair is given. Such a sequence has to contain a specific sequence pattern,
facilitating the switch from one structure to the other. Various studies are trying to give criteria
for riboswitch sequences. \cite{Clote:07} considers the Boltzmann ensemble of a sequence with base pair
filtration. If there are multiple clusters of structures which have high density and a fixed base pair
distance to the mfe-configuration, then alternative structures are predicted. However, the analysis
specifies the existence of structure clusters relative to the mfe-structure and it is not clear if they
can switch. The ranking $(r_R, r_S)$ in Figure~\ref{F:ranking} allows to identify native riboswitch \emph{sequences} from Boltzmann
sampled sequences of the native riboswitch structure pair. Thus, for riboswitches both: the phenotype pair
and the genotype are distinguished. The alternative structures of riboswitches have a dense set of
bicompatible sequences and the native sequence assumes a particularly low energy for the alternative
configurations. The latter is displayed in Figure~\ref{F:ranking}. Boltzmann sampled sequences for
riboswitch structure pairs are distinctively different from the native sequence. Thus the design of
riboswitch sequences involves two types of data: the structure pair, as well as the sequence simultaneously.




\begin{backmatter}

\section*{Competing interests}
  The authors declare that they have no competing interests.


\section*{Acknowledgements}
We want to thank Andrei Bura, Qijun He and Thomas Li for their input and discussion on this manuscript. 
We gratefully acknowledge the help of Henning Mortveit, Madhav Marathe and Reza Rezazadegan for their computational supports.




\newcommand{\BMCxmlcomment}[1]{}

\BMCxmlcomment{

<refgrp>

<bibl id="B1">
  <title><p>RNA folding and combinatory landscapes</p></title>
  <aug>
    <au><snm>Fontana</snm><fnm>W.</fnm></au>
    <au><snm>F.</snm><fnm>SP</fnm></au>
    <au><snm>Bornberg Bauer</snm><fnm>EG</fnm></au>
    <au><snm>Griesmacher</snm><fnm>T</fnm></au>
    <au><snm>L.</snm><fnm>HI</fnm></au>
    <au><snm>Tacker</snm><fnm>M.</fnm></au>
    <au><snm>Tarazona</snm><fnm>P.</fnm></au>
    <au><snm>Weinberger</snm><fnm>E. D.</fnm></au>
    <au><snm>Schuster</snm><fnm>P.</fnm></au>
  </aug>
  <source>Phys. .Rev. E</source>
  <pubdate>1993</pubdate>
  <volume>47</volume>
  <fpage>2083</fpage>
  <lpage>-2099</lpage>
</bibl>

<bibl id="B2">
  <title><p>Analysis of {RNA} sequence structure maps by exhaustive enumeration
  {I}. Neutral networks</p></title>
  <aug>
    <au><snm>Gr{\"u}ner</snm><fnm>W.</fnm></au>
    <au><snm>Giegerich</snm><fnm>R.</fnm></au>
    <au><snm>Strothmann</snm><fnm>D.</fnm></au>
    <au><snm>Reidys</snm><fnm>C.</fnm></au>
    <au><snm>Weber</snm><fnm>J.</fnm></au>
    <au><snm>Hofacker</snm><fnm>I. L.</fnm></au>
    <au><snm>Stadler</snm><fnm>P. F.</fnm></au>
    <au><snm>Schuster</snm><fnm>P.</fnm></au>
  </aug>
  <source>Monatsh.\ Chem.</source>
  <pubdate>1996</pubdate>
  <volume>127</volume>
  <fpage>355</fpage>
  <lpage>-374</lpage>
</bibl>

<bibl id="B3">
  <title><p>Analysis of {RNA} sequence structure maps by exhaustive enumeration
  {II.} Structures of neutral networks and shape space covering</p></title>
  <aug>
    <au><snm>Gr{\"u}ner</snm><fnm>W.</fnm></au>
    <au><snm>Giegerich</snm><fnm>R.</fnm></au>
    <au><snm>Strothmann</snm><fnm>D.</fnm></au>
    <au><snm>Reidys</snm><fnm>C.</fnm></au>
    <au><snm>Weber</snm><fnm>J.</fnm></au>
    <au><snm>Hofacker</snm><fnm>I. L.</fnm></au>
    <au><snm>Stadler</snm><fnm>P. F.</fnm></au>
    <au><snm>Schuster</snm><fnm>P.</fnm></au>
  </aug>
  <source>Monatsh.\ Chem.</source>
  <pubdate>1996</pubdate>
  <volume>127</volume>
  <fpage>375</fpage>
  <lpage>-389</lpage>
</bibl>

<bibl id="B4">
  <title><p>From sequences to shapes and back: a case study in {RNA} secondary
  structures</p></title>
  <aug>
    <au><snm>Schuster</snm><fnm>P</fnm></au>
    <au><snm>Fontana</snm><fnm>W.</fnm></au>
    <au><snm>Stadler</snm><fnm>P. F.</fnm></au>
    <au><snm>Hofacker</snm><fnm>I. L.</fnm></au>
  </aug>
  <source>Proc Biol Sci.</source>
  <pubdate>1994</pubdate>
  <volume>255</volume>
  <issue>1344</issue>
  <fpage>279</fpage>
  <lpage>84</lpage>
</bibl>

<bibl id="B5">
  <title><p>Generic properties of combinatory maps: Neural networks of {RNA}
  secondary structures+</p></title>
  <aug>
    <au><snm>Reidys</snm><fnm>C. M.</fnm></au>
    <au><snm>Stadler</snm><fnm>P. F.</fnm></au>
    <au><snm>Schuster</snm><fnm>P.</fnm></au>
  </aug>
  <source>Bull. Math. Biol.</source>
  <pubdate>1997</pubdate>
  <volume>59</volume>
  <fpage>339</fpage>
  <lpage>397</lpage>
</bibl>

<bibl id="B6">
  <title><p>Continuity in evolution: on the nature of transitions</p></title>
  <aug>
    <au><snm>Fontana</snm><fnm>W.</fnm></au>
    <au><snm>Schuster</snm><fnm>P.</fnm></au>
  </aug>
  <source>Science</source>
  <pubdate>1998</pubdate>
  <volume>280</volume>
  <issue>5368</issue>
  <fpage>1451</fpage>
  <lpage>-5</lpage>
</bibl>

<bibl id="B7">
  <title><p>Genotypes with phenotypes: adventures in an {RNA} toy
  world</p></title>
  <aug>
    <au><snm>Schuster</snm><fnm>P.</fnm></au>
  </aug>
  <source>Biophys. Chem.</source>
  <pubdate>1997</pubdate>
  <volume>66</volume>
  <issue>2-3</issue>
  <fpage>75</fpage>
  <lpage>110</lpage>
</bibl>

<bibl id="B8">
  <title><p>Evolutionary rate at the molecular level</p></title>
  <aug>
    <au><snm>Kimura</snm><fnm>M.</fnm></au>
  </aug>
  <source>Nature</source>
  <pubdate>1968</pubdate>
  <volume>217</volume>
  <fpage>624</fpage>
  <lpage>-626</lpage>
</bibl>

<bibl id="B9">
  <title><p>Evolutionary Dynamics and Optimization: Neutral Networks as
  Model-Landscapes for {RNA} Secondary-Structure Folding-Landscapes</p></title>
  <aug>
    <au><snm>Forst</snm><fnm>C. V.</fnm></au>
    <au><snm>Reidys</snm><fnm>C. M.</fnm></au>
    <au><snm>Weber</snm><fnm>J.</fnm></au>
  </aug>
  <source>ECAL</source>
  <pubdate>1995</pubdate>
  <fpage>128</fpage>
  <lpage>-147</lpage>
</bibl>

<bibl id="B10">
  <title><p>Riboswitches: structures and mechanisms</p></title>
  <aug>
    <au><snm>Garst</snm><fnm>A. D.</fnm></au>
    <au><snm>Edwards</snm><fnm>A. L.</fnm></au>
    <au></au>
    <au><snm>Batey</snm><fnm>R. T.</fnm></au>
  </aug>
  <source>Cold Spring Harb Perspect Biol,</source>
  <pubdate>2011</pubdate>
  <volume>3</volume>
  <issue>6</issue>
</bibl>

<bibl id="B11">
  <title><p>A decade of riboswitches</p></title>
  <aug>
    <au><snm>Serganov</snm><fnm>A.</fnm></au>
    <au><snm>Nudler</snm><fnm>E.</fnm></au>
  </aug>
  <source>Cell</source>
  <pubdate>2011</pubdate>
  <volume>152</volume>
  <issue>1-2</issue>
  <fpage>17</fpage>
  <lpage>-24</lpage>
</bibl>

<bibl id="B12">
  <title><p>Ligand-induced folding of the adenosine deaminase a-riboswitch and
  implications on riboswitch translational control</p></title>
  <aug>
    <au><snm>Rieder</snm><fnm>R</fnm></au>
    <au><snm>Lang</snm><fnm>K</fnm></au>
    <au><snm>Graber</snm><fnm>D</fnm></au>
    <au><snm>Micura</snm><fnm>R</fnm></au>
  </aug>
  <source>Chem. Bio. Chem.</source>
  <pubdate>2007</pubdate>
  <volume>8</volume>
  <issue>896--902</issue>
</bibl>

<bibl id="B13">
  <title><p>Design of multistable {RNA} molecules</p></title>
  <aug>
    <au><snm>Flamm</snm><fnm>C.</fnm></au>
    <au><snm>Hofacker</snm><fnm>I. L.</fnm></au>
    <au><snm>Maurer Stroh</snm><fnm>S.</fnm></au>
    <au><snm>Stadler</snm><fnm>P. F.</fnm></au>
    <au><snm>Zehl</snm><fnm>M.</fnm></au>
  </aug>
  <source>RNA</source>
  <pubdate>2001</pubdate>
  <volume>7</volume>
  <issue>2</issue>
  <fpage>254</fpage>
  <lpage>-265</lpage>
</bibl>

<bibl id="B14">
  <title><p>Frnakenstein: multiple target inverse {RNA} folding</p></title>
  <aug>
    <au><snm>Lyngs{\o}</snm><fnm>RB</fnm></au>
    <au><snm>Anderson</snm><fnm>JW</fnm></au>
    <au><snm>Sizikova</snm><fnm>E</fnm></au>
    <au><snm>Badugu</snm><fnm>A</fnm></au>
    <au><snm>Hyland</snm><fnm>T</fnm></au>
    <au><snm>Hein</snm><fnm>J</fnm></au>
  </aug>
  <source>BMC bioinformatics</source>
  <publisher>BioMed Central</publisher>
  <pubdate>2012</pubdate>
  <volume>13</volume>
  <issue>1</issue>
  <fpage>260</fpage>
</bibl>

<bibl id="B15">
  <title><p>Multi-objective optimization for {RNA} design with multiple target
  secondary structures</p></title>
  <aug>
    <au><snm>Taneda</snm><fnm>A</fnm></au>
  </aug>
  <source>BMC bioinformatics</source>
  <publisher>BioMed Central</publisher>
  <pubdate>2015</pubdate>
  <volume>16</volume>
  <issue>1</issue>
  <fpage>280</fpage>
</bibl>

<bibl id="B16">
  <title><p>{RNAblueprint}: flexible multiple target nucleic acid sequence
  design</p></title>
  <aug>
    <au><snm>Hammer</snm><fnm>S</fnm></au>
    <au><snm>Tschiatschek</snm><fnm>B</fnm></au>
    <au><snm>Flamm</snm><fnm>C</fnm></au>
    <au><snm>Hofacker</snm><fnm>IL</fnm></au>
    <au><snm>Findei{\ss}</snm><fnm>S</fnm></au>
  </aug>
  <source>Bioinformatics</source>
  <publisher>Oxford University Press</publisher>
  <pubdate>2017</pubdate>
  <volume>33</volume>
  <issue>18</issue>
  <fpage>2850</fpage>
  <lpage>-2858</lpage>
</bibl>

<bibl id="B17">
  <title><p>Fixed-parameter tractable sampling for {RNA} design with multiple
  target structures</p></title>
  <aug>
    <au><snm>Hammer</snm><fnm>S</fnm></au>
    <au><snm>Wang</snm><fnm>W</fnm></au>
    <au><snm>Will</snm><fnm>S</fnm></au>
    <au><snm>Ponty</snm><fnm>Y</fnm></au>
  </aug>
  <source>BMC bioinformatics</source>
  <publisher>BioMed Central</publisher>
  <pubdate>2019</pubdate>
  <volume>20</volume>
  <issue>1</issue>
  <fpage>209</fpage>
</bibl>

<bibl id="B18">
  <title><p>Fast Folding and Comparison of {RNA} Secondary
  Structures</p></title>
  <aug>
    <au><snm>Hofacker</snm><fnm>IL</fnm></au>
    <au><snm>Fontana</snm><fnm>W</fnm></au>
    <au><snm>Stadler</snm><fnm>PF</fnm></au>
    <au><snm>Bonhoeffer</snm><fnm>LS</fnm></au>
    <au><snm>Tacker</snm><fnm>M</fnm></au>
    <au><snm>Schuster</snm><fnm>P</fnm></au>
  </aug>
  <source>Monatsh.\ Chem.</source>
  <pubdate>1994</pubdate>
  <volume>125</volume>
  <fpage>167</fpage>
  <lpage>188</lpage>
</bibl>

<bibl id="B19">
  <title><p>A global sampling approach to designing and reengineering {RNA}
  secondary structures</p></title>
  <aug>
    <au><snm>Levin</snm><fnm>A.</fnm></au>
    <au><snm>Lis</snm><fnm>M.</fnm></au>
    <au><snm>Ponty</snm><fnm>Y.</fnm></au>
    <au><snm>O'Donnell</snm><fnm>C. W.</fnm></au>
    <au><snm>Devadas</snm><fnm>S.</fnm></au>
    <au><snm>Berger</snm><fnm>B.</fnm></au>
    <au><snm>Waldisp{\"u}hl</snm><fnm>J.</fnm></au>
  </aug>
  <source>Nucl. Acids Res.</source>
  <pubdate>2012</pubdate>
  <volume>40</volume>
  <issue>20</issue>
  <fpage>10041</fpage>
  <lpage>10052</lpage>
</bibl>

<bibl id="B20">
  <title><p>A weighted sampling algorithm for the design of {RNA} sequences
  with targeted secondary structure and nucleotide distribution</p></title>
  <aug>
    <au><snm>Reinharz</snm><fnm>V.</fnm></au>
    <au><snm>Ponty</snm><fnm>Y.</fnm></au>
    <au><snm>Waldisp{\"u}hl</snm><fnm>J.</fnm></au>
  </aug>
  <source>Bioinformatics</source>
  <pubdate>2013</pubdate>
  <volume>29</volume>
  <issue>13</issue>
  <fpage>308</fpage>
  <lpage>315</lpage>
</bibl>

<bibl id="B21">
  <title><p>{RNAdualPF}: software to compute the dual partition function with
  sample applications in molecular evolution theory.</p></title>
  <aug>
    <au><snm>Garcia Martin</snm><fnm>J. A.</fnm></au>
    <au><snm>Bayegan</snm><fnm>A. H.</fnm></au>
    <au><snm>Dotu</snm><fnm>I.</fnm></au>
    <au><snm>Clote</snm><fnm>P.</fnm></au>
  </aug>
  <source>BMC Bioinformatics</source>
  <pubdate>2016</pubdate>
  <volume>17</volume>
  <issue>1</issue>
  <fpage>424</fpage>
</bibl>

<bibl id="B22">
  <title><p>Sequence-structure relations of biopolymers.</p></title>
  <aug>
    <au><snm>Barrett</snm><fnm>C.</fnm></au>
    <au><snm>Huang</snm><fnm>F.</fnm></au>
    <au><snm>Reidys</snm><fnm>C. M.</fnm></au>
  </aug>
  <source>Bioinformatics</source>
  <pubdate>2017</pubdate>
  <volume>33</volume>
  <issue>3</issue>
  <fpage>382</fpage>
  <lpage>389</lpage>
</bibl>

<bibl id="B23">
  <title><p>The equilibrium partition function and base pair binding
  probabilities for {RNA} secondary structure</p></title>
  <aug>
    <au><snm>McCaskill</snm><fnm>J. S.</fnm></au>
  </aug>
  <source>Biopolymers</source>
  <pubdate>1990</pubdate>
  <volume>29</volume>
  <fpage>1105</fpage>
  <lpage>1119</lpage>
</bibl>

<bibl id="B24">
  <title><p>A linear-time algorithm for finding tree-decompositions of small
  treewidth</p></title>
  <aug>
    <au><snm>Bodlaender</snm><fnm>HL</fnm></au>
  </aug>
  <source>SIAM Journal on computing</source>
  <publisher>SIAM</publisher>
  <pubdate>1996</pubdate>
  <volume>25</volume>
  <issue>6</issue>
  <fpage>1305</fpage>
  <lpage>-1317</lpage>
</bibl>

<bibl id="B25">
  <title><p>Expanded sequence dependence of thermodynamic parameters improves
  prediction of \textsc{RNA} secondary structure</p></title>
  <aug>
    <au><snm>Mathews</snm><fnm>D.</fnm></au>
    <au><snm>Sabina</snm><fnm>J.</fnm></au>
    <au><snm>Zuker</snm><fnm>M.</fnm></au>
    <au><snm>Turner</snm><fnm>D.H.</fnm></au>
  </aug>
  <source>J. Mol. Biol.</source>
  <pubdate>1999</pubdate>
  <volume>288</volume>
  <fpage>911</fpage>
  <lpage>940</lpage>
</bibl>

<bibl id="B26">
  <title><p>Loop Homology of Bi-secondary Structures</p></title>
  <aug>
    <au><snm>Bura</snm><fnm>A. C.</fnm></au>
    <au><snm>He</snm><fnm>Q.</fnm></au>
    <au><snm>Reidys</snm><fnm>C. M.</fnm></au>
  </aug>
  <source>arXiv:1904.02041</source>
  <pubdate>2019</pubdate>
</bibl>

<bibl id="B27">
  <title><p>Traveling salesman problem heuristics: Leading methods,
  implementations and latest advances</p></title>
  <aug>
    <au><snm>Rego</snm><fnm>C</fnm></au>
    <au><snm>Gamboa</snm><fnm>D</fnm></au>
    <au><snm>Glover</snm><fnm>F</fnm></au>
    <au><snm>Osterman</snm><fnm>C</fnm></au>
  </aug>
  <source>European Journal of Operational Research</source>
  <publisher>Elsevier</publisher>
  <pubdate>2011</pubdate>
  <volume>211</volume>
  <issue>3</issue>
  <fpage>427</fpage>
  <lpage>-441</lpage>
</bibl>

<bibl id="B28">
  <title><p>Phenotypic Switching Can Speed up Microbial Evolution</p></title>
  <aug>
    <au><snm>Tadrowski</snm><fnm>AC</fnm></au>
    <au><snm>Evans</snm><fnm>MR</fnm></au>
    <au><snm>Waclaw</snm><fnm>B</fnm></au>
  </aug>
  <source>Scientific reports</source>
  <publisher>Nature Publishing Group</publisher>
  <pubdate>2018</pubdate>
  <volume>8</volume>
  <issue>1</issue>
  <fpage>8941</fpage>
</bibl>

<bibl id="B29">
  <title><p>Strong phenotypic plasticity limits potential for evolutionary
  responses to climate change</p></title>
  <aug>
    <au><snm>Oostra</snm><fnm>V</fnm></au>
    <au><snm>Saastamoinen</snm><fnm>M</fnm></au>
    <au><snm>Zwaan</snm><fnm>BJ</fnm></au>
    <au><snm>Wheat</snm><fnm>CW</fnm></au>
  </aug>
  <source>Nature communications</source>
  <publisher>Nature Publishing Group</publisher>
  <pubdate>2018</pubdate>
  <volume>9</volume>
  <issue>1</issue>
  <fpage>1005</fpage>
</bibl>

<bibl id="B30">
  <title><p>Incorporating chemical modification constraints into a dynamic
  programming algorithm for prediction of {RNA} secondary structure</p></title>
  <aug>
    <au><snm>Mathews</snm><fnm>D</fnm></au>
    <au><snm>Disney</snm><fnm>M</fnm></au>
    <au><snm>Childs</snm><fnm>J</fnm></au>
    <au><snm>Schroeder</snm><fnm>S</fnm></au>
    <au><snm>Zuker</snm><fnm>M</fnm></au>
    <au><snm>Turner</snm><fnm>D.</fnm></au>
  </aug>
  <source>Proc Natl Acad Sci</source>
  <pubdate>2004</pubdate>
  <volume>101</volume>
  <fpage>7287</fpage>
  <lpage>7292</lpage>
</bibl>

<bibl id="B31">
  <title><p>Fatgraph models of proteins</p></title>
  <aug>
    <au><snm>Penner</snm><fnm>R. C.</fnm></au>
    <au><snm>Knudsen</snm><fnm>M</fnm></au>
    <au><snm>Wiuf</snm><fnm>C</fnm></au>
    <au><snm>Andersen</snm><fnm>JE</fnm></au>
  </aug>
  <source>Comm.\ Pure Appl.\ Math.</source>
  <pubdate>2010</pubdate>
  <volume>63</volume>
  <fpage>1249</fpage>
  <lpage>1297</lpage>
</bibl>

<bibl id="B32">
  <title><p>A statistical sampling algorithm for {RNA} secondary structure
  prediction</p></title>
  <aug>
    <au><snm>Ding</snm><fnm>Y.</fnm></au>
    <au><snm>Lawrence</snm><fnm>C. E.</fnm></au>
  </aug>
  <source>Nucleic Acids Res.</source>
  <pubdate>2003</pubdate>
  <volume>31</volume>
  <fpage>7280</fpage>
  <lpage>7301</lpage>
</bibl>

<bibl id="B33">
  <title><p>Algebraic Topology: An Introduction</p></title>
  <aug>
    <au><snm>Massey</snm><fnm>WS</fnm></au>
  </aug>
  <publisher>Springer-Veriag, New York</publisher>
  <pubdate>1967</pubdate>
</bibl>

<bibl id="B34">
  <title><p>Structure of a natural guanine-responsive riboswitch complexed with
  the metabolite hypoxanthine</p></title>
  <aug>
    <au><snm>Batey</snm><fnm>RT</fnm></au>
    <au><snm>Gilbert</snm><fnm>SD</fnm></au>
    <au><snm>Montange</snm><fnm>RK</fnm></au>
  </aug>
  <source>Nature</source>
  <pubdate>2004</pubdate>
  <volume>432</volume>
  <issue>7015</issue>
  <fpage>411</fpage>
  <lpage>-415</lpage>
</bibl>

<bibl id="B35">
  <title><p>Structure and mechanism of a metal-sensing regulatory
  {RNA}</p></title>
  <aug>
    <au><snm>Dann</snm><fnm>CE</fnm></au>
    <au><snm>Wakeman</snm><fnm>CA</fnm></au>
    <au><snm>Sieling</snm><fnm>CL</fnm></au>
    <au><snm>Baker</snm><fnm>SC</fnm></au>
    <au><snm>Irnov</snm><fnm>I</fnm></au>
    <au><snm>Winkler</snm><fnm>WC</fnm></au>
  </aug>
  <source>Cell</source>
  <pubdate>2007</pubdate>
  <volume>130</volume>
  <issue>5</issue>
  <fpage>878</fpage>
  <lpage>-892</lpage>
</bibl>

<bibl id="B36">
  <title><p>An mRNA structure in bacteria that controls gene expression by
  binding lysine</p></title>
  <aug>
    <au><snm>Sudarsan</snm><fnm>N</fnm></au>
    <au><snm>Wickiser</snm><fnm>JK</fnm></au>
    <au><snm>Nakamura</snm><fnm>S</fnm></au>
    <au><snm>Ebert</snm><fnm>MS</fnm></au>
    <au><snm>Breaker</snm><fnm>RR</fnm></au>
  </aug>
  <source>Genes Dev.</source>
  <pubdate>2003</pubdate>
  <volume>17</volume>
  <issue>21</issue>
  <fpage>2688</fpage>
  <lpage>-2697</lpage>
</bibl>

<bibl id="B37">
  <title><p>A stress-responsive RNA switch regulates {VEGFA}
  expression</p></title>
  <aug>
    <au><snm>Ray</snm><fnm>PS</fnm></au>
    <au><snm>Jia</snm><fnm>J</fnm></au>
    <au><snm>Yao</snm><fnm>P</fnm></au>
    <au><snm>Mithu</snm><fnm>M</fnm></au>
    <au><snm>Hatzoglou</snm><fnm>M</fnm></au>
    <au><snm>Fox</snm><fnm>PL</fnm></au>
  </aug>
  <source>Nature</source>
  <pubdate>2009</pubdate>
  <volume>457</volume>
  <issue>7231</issue>
  <fpage>915</fpage>
  <lpage>-919</lpage>
</bibl>

<bibl id="B38">
  <title><p>Natural variability in {S}-adenosylmethionine {(SAM)}-dependent
  riboswitches: {S}-box elements in bacillus subtilis exhibit differential
  sensitivity to {SAM} in vivo and in vitro</p></title>
  <aug>
    <au><snm>Tomsic</snm><fnm>J</fnm></au>
    <au><snm>McDaniel</snm><fnm>BA</fnm></au>
    <au><snm>Grundy</snm><fnm>FJ</fnm></au>
    <au><snm>Henkin</snm><fnm>TM</fnm></au>
  </aug>
  <source>J. Bacteriol</source>
  <pubdate>2008</pubdate>
  <volume>190</volume>
  <issue>3</issue>
  <fpage>823</fpage>
  <lpage>-833</lpage>
</bibl>

<bibl id="B39">
  <title><p>Central and local limit theorems for RNA structures</p></title>
  <aug>
    <au><snm>Jin</snm><fnm>EY</fnm></au>
    <au><snm>Reidys</snm><fnm>C. M.</fnm></au>
  </aug>
  <source>J. Theo. Biol.</source>
  <pubdate>2008</pubdate>
  <volume>250</volume>
  <issue>3</issue>
  <fpage>547</fpage>
  <lpage>-559</lpage>
</bibl>

<bibl id="B40">
  <title><p>{INFO-RNA}--a fast approach to inverse RNA folding</p></title>
  <aug>
    <au><snm>Busch</snm><fnm>A.</fnm></au>
    <au><snm>Backofen</snm><fnm>R.</fnm></au>
  </aug>
  <source>Bioinformatics</source>
  <pubdate>2006</pubdate>
  <volume>22</volume>
  <issue>15</issue>
  <fpage>1823</fpage>
  <lpage>31</lpage>
</bibl>

<bibl id="B41">
  <title><p>Boltzmann probability of {RNA} structural neighbors and riboswitch
  detection</p></title>
  <aug>
    <au><snm>Freyhult</snm><fnm>E.</fnm></au>
    <au><snm>Moulton</snm><fnm>V.</fnm></au>
    <au><snm>Clote</snm><fnm>P.</fnm></au>
  </aug>
  <source>Bioinformatics,</source>
  <pubdate>2007</pubdate>
  <volume>23</volume>
  <issue>16</issue>
  <fpage>2054</fpage>
  <lpage>-2062</lpage>
</bibl>

</refgrp>
} 

\end{backmatter}
\end{document}